\documentclass[namedreferences,hyperref,optionalrh]{spr-sola}

\ifdefined\NAT@force@numbers
  \NAT@force@numbers
\fi
\usepackage{graphicx}       
\usepackage{color}     
                       
\usepackage[normalem]{ulem}
\usepackage{amsmath}
\usepackage{placeins}
\usepackage{afterpage}

\usepackage[dvipsnames]{xcolor}

\chardef\us=`\_
\definecolor{ps}{RGB}{255, 0, 0}   
\newcommand{\authA}[1]{\textcolor{ps}{[ps]}}
\newcommand{\spc}{\texttt{SPACE}}

\begin{document}

\begin{frontmatter}
\title{{SPACE-SUIT:An Artificial Intelligence Based Chromospheric Feature Extractor and Classifier for SUIT}}
\author[addressref={aff1,aff2},corref,email={pranavaseth@gmail.com}]{\fnm{Pranava }~\snm{Seth}\orcid{0009-0001-7184-6797}}
\author[addressref={aff3,aff4},corref,email={vishal@lmsal.com}]
{\fnm{Vishal}~\snm{Upendran}\orcid{0000-0002-9253-6093}}
\author[addressref={aff5,aff2},corref,email={megha.anand@manipal.edu}]{\fnm{Megha}~\snm{Anand}\orcid{0000-0002-8457-9771}}
\author[addressref={aff2,aff6},corref,email={janmejoy.sarkar@iucaa.in}]
{\fnm{Janmejoy}~\snm{Sarkar}\orcid{0000-0002-8560-318X}}
\author[addressref={aff2},corref,email={soumyaroy@iucaa.in}]{\fnm{Soumya}~\snm{Roy}\orcid{0000-0003-2215-7810}}
\author[addressref={aff9},corref,email={pdsnchaki@gmail.com}]{\fnm{Priyadarshan}~\snm{Chaki}}
\author[addressref={aff7},corref,email={pratyaychowdhury725@gmail.com}]{\fnm{Pratyay}~\snm{Chowdhury}}
\author[addressref={aff8},corref,email={borishan.gh@gmail.com
}]{\fnm{Borishan}~\snm{Ghosh}}
\author[addressref={aff2},corref,email={durgesh@iucaa.in}]{\fnm{Durgesh}~\snm{Tripathi}\orcid{0000-0003-1689-6254}}
\address[id=aff1]{Thapar Institute of Engineering and Technology, Patiala, India - 147004}
\address[id=aff2]{Inter-University Centre for Astronomy and Astrophysics, Pune, India - 411007}
\address[id=aff3]{SETI Institute, Mountain View, CA, USA - 94043}
\address[id=aff4]{Lockheed Martin Solar and Astrophysics Laboratory, Palo Alto, CA, USA - 94304}
\address[id=aff5]{Manipal Centre for Natural Sciences, Manipal Academy of Higher Education, Manipal, Karnataka, India - 576104}
\address[id=aff6]{Department of Physics, Tezpur University, Napaam, Tezpur, Assam, India- 784028}
\address[id=aff9]{Ramakrishna Mission Residential College, Narendrapur,Kolkata, Rajpur Sonarpur, West Bengal, India - 700103}
\address[id=aff7]{Indian Association for the Cultivation of Science,Jadavpur, Kolkata, West Bengal, India - 700032}
\address[id=aff8]{Hansraj College, University of Delhi, Delhi, India - 110007}

\runningauthor{Seth et al.}
\runningtitle{SPACE-SUIT}

\begin{abstract}
The Solar Ultraviolet Imaging Telescope (SUIT) onboard Aditya-L1 is an imager that observes the solar photosphere and chromosphere through observations in the wavelength range of 200\,--400\, nm. A comprehensive understanding of the plasma and thermodynamic properties of chromospheric and photospheric morphological structures requires a large sample statistical study of these regions, necessitating the development of automatic feature detection methods. To this end, we develop the feature detection algorithm \spc-\texttt{SUIT}: \underline{S}olar \underline{P}henomena \underline{A}nalysis and \underline{C}lassification using \underline{E}nhanced vision techniques for SUIT, to detect and classify the solar chromospheric features to be observed from SUIT’s Mg~\rm{II}~k filter. Specifically, we target plage regions, sunspots, filaments, and off-limb structures for detection using this algorithm. SPACE uses You Only Look Once (YOLO), a neural network-based model to identify regions of interest. We train and validate SPACE using mock-SUIT images developed from Interface Region Imaging Spectrometer (IRIS) full-disk mosaic images in Mg~\rm{II}~k line, while we also perform detection on Level-1 SUIT data. {\spc} achieves a precision of    $\approx0.788$, recall of $\approx0.863$ and a MAP of $\approx0.874$  on the validation mock SUIT FITS dataset. Since our dataset is manually labeled, we perform `self-validation' on the identified regions by defining statistical measures and Tamura features on the ground truth and predicted bounding boxes. We find the distributions of entropy, contrast, dissimilarity, and energy to show differences for the features in consideration. We find these differences to be captured qualitatively by the detected regions predicted by {\spc}. Furthermore, we find these differences to also be qualitatively captured by the observed SUIT images, reflecting validation in the absence of a labeled ground truth. This work hence not only develops a chromospheric feature extractor, but it also demonstrates the effectiveness of statistical metrics and Tamura features in differentiating chromospheric features of interest, providing independent validation measures for any future detection and validation scheme.
\end{abstract}

\end{frontmatter}
\section{Introduction}
\label{sec:introduction} 
The lower solar atmosphere illustrates a complex interplay of fluid and magnetic processes that modulate the properties of the upper atmosphere and the heliosphere. Especially, the chromosphere is a complex region where magnetic and fluid effects play almost equal roles, manifesting several morphological structures like plage regions, filaments, sunspots, fibrils, and spicules. These structures are known to be coupled across the atmosphere, strongly influencing the dynamics across the atmosphere.  Hence, understanding the properties and formation of these structures is crucial in understanding the dynamics of the solar atmosphere.

Aditya-L1 \citep{AL1} is India's first space-based solar observatory. It was launched on 2 September 2023, and inserted into an L1-halo orbit around the Sun on 6 January 2024. The observatory has several remote-sensing and in-situ measurement instruments onboard, including spectrometers, plasma analyzers, and spectral imagers. The Solar Ultraviolet Imaging Telescope (SUIT) \citep{Tripathi_2017} is a spectral photoimager onboard Aditya-L1, observing the solar atmosphere in the 2000\,-- 4000\, {\AA} wavelength range in eleven spectral bandpasses probing the photosphere and chromosphere. These spectral bandpasses contain 8 narrow bands and 3 broad bands, with a plate scale of 0.7''/px. SUIT produces routine full-disc images in the NB3 filter, centered around the Mg II k line at $\approx2796$~{\AA}. Extensive statistical studies of chromospheric features of interest demand automated feature detection and extraction methods operating on the NB3 passband. Hence, it is hence imperative to develop algorithms to detect chromospheric morphological features observed by SUIT, which will be made available to the community as a Level 2+ data product by the team. 

Feature extraction, detection and segmentation are well-known problems in heliophysics, solved either using empirical or learning-based methods. For instance, sunspot identification has been a classic problem, with several efforts on automating the process~\citep{sunspot_colak_2008SoPh..248..277C,sunspot_goel_2014SoPh..289.1413G,sreejith_sunspot_2016SoPh..291...41P,sunspot_zhao_2016,Sunspots__Identification_Through_Mathematical_Morphology,Automatic_detection_of_sunspots}. Similarly, automated filament detection is a classic problem, with several efforts in the past~\citep{Automatic_Solar_Filament,Automatic_Solar_Filament_Segmentation_and_Characterization,Toward_Filament_Segmentation,Solar_Filament_Segmentation,Filament_Detection,filament_andrea_2024A&A...686A.213D}, along with detection of plages~\citep{plage_benkhalil_2006,Solar_plages_det}. Most of these systems are demonstrated with either (i). A specific feature of interest, or (ii). A specific telescope/set of telescopes of interest. More precisely, each telescope in consideration has observed the Sun in different bandpasses, has different instrument characteristics, and hence would have different features of interest. 

In this article, we describe the development of a feature detection system for SUIT, called {\spc}:  \underline{S}olar \underline{P}henomena \underline{A}nalysis and \underline{C}lassification using \underline{E}nhanced vision techniques. This is a supervised deep-learning algorithm and is based on the visual detection model YOLO~\citep{v9}. With {\spc}, we are specifically interested in the following features from SUIT: (i). Plages, (ii). Filaments, (iii). Sunspots, and (iv). Off-limb structures, and deploy the algorithm for the NB3 passband of SUIT. 

In general, the evaluation of any feature identification algorithm critically depends on a well-defined ground truth. The definition of ground truth, in turn, depends on the specific feature identification method, which may be manual or automatic. Hence, it is imperative we evaluate the proposed algorithm through a framework that is more resilient to issues in the definition of ground truth. Hence, in this work, we propose a framework using statistical measures and Tamura features to evaluate both the ground truth and predictions, presenting measures that may be used for auto-validation of feature detection methods. Previous studies have demonstrated the utility of statistical moments, such as skewness and kurtosis, in effectively analyzing solar flare activities and revealing pre-flare signatures~\citep{Statistical_Moments_of_Active_Region_Images_During_Solar_Flares}. Similarly, texture-based features, including Tamura directionality and contrast, which have been widely utilized for solar image analysis, have shown significant potential, particularly for feature recognition and tracking~\citep{Region_based_querying_of_solar_data_using_descriptor_signatures, large_scale_solar_image_dataset_with_labeled_event_regions, Massive_Labeled_Solar_Image_Data_Benchmarks_for_Automated_Feature_Recognition, Iterative_refinement_of_multiple_targets_tracking_of_solar_events, Improving_the_Functionality_of_Tamura_Directionality_on_Solar_Images, Tracking_Solar_Phenomena_from_the_SDO}.
We train and evaluate {\spc} on mock-SUIT images, and perform a zero-shot prediction on the SUIT observations.

In the following sections, we first describe the development of mock SUIT images in Section~\ref{sec:mock_suit_gen}, and describe the data processing in Section~\ref{sec:data_preprocessing}. In Section~\ref{sec:MAE}, we present the  automatic detection algorithm. In Section~\ref{sec:STF}, we present a self-validation scheme using statistical and Tamura information, while in Section~\ref{sec:results}, we also present all the results and a definition of the compiled catalog.

\section{Dataset Preparation}
\label{sec:data}

\subsection{Mock SUIT Image Generation}
\label{sec:mock_suit_gen}
\begin{figure}[!ht]
    \centering
    \includegraphics[width=0.8\textwidth]{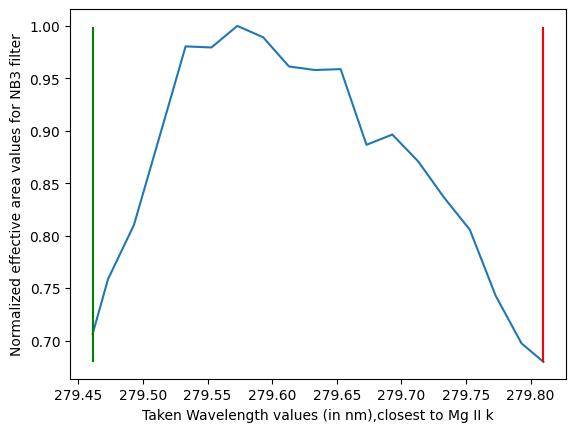}
    \caption{Effective Area of NB3 filter of SUIT normalized by its peak, coinciding with the wavelength range IRIS Mg~\rm{II}~k mosaics.
    The \textit{green} and \textit{red} vertical lines specify lower and higher wavelength limits recorded by IRIS, respectively. }
    \label{fig:suit_effective_area}
\end{figure}
{\spc} is a supervised learning algorithm, that needs pairs of {input images, bounding boxes}. These input images, on deployment of the workflow, would be the SUIT images. As of this manuscript, SUIT has not generated enough data for developing a supervised detection model. Hence, we develop a catalogue of `mock' SUIT images using data recorded by the Interface Region Imaging Spectrograph (IRIS)~\citep{iris}.

IRIS contains a spectrograph and a slit-jaw imager, and performs observations in near ultraviolet (NUV) and far ultraviolet (FUV) regimes. The spectroscopy is performed in three windows, one in near ultraviolet (NUV) from $2782.7$ to $2851.1$~{\AA}, and two in the far ultraviolet (FUV), from 1332 to 1358~{\AA} (FUV 1), and from 1389 to 1407~{\AA} (FUV 2).  IRIS periodically takes a full-disk mosaic in the 6 spectral windows surrounding the strongest spectral lines, by combining rasters taken at different pointings covering the full disc. The observation program covers $\approx185$ pointings across $\approx18$ hours, by constructing a $64$-step raster with step size of $\approx$ 2 arcsec, at an exposure time of 1-2 seconds, with a spectral resolution of 0.035~{\AA}. These rasters are available on the IRIS website~\footnote{\href{https://iris.lmsal.com/mosaic\_allin1.html}{iris.lmsal.com/mosaic\_allin1.html}}. An example set of IRIS mosaics at 2796~{\AA} is shown in  Figure~\ref{fig:suitmock1}a.

We use $\approx10$ years (30-09-2013 -- 23-07-2023) of Mg II k full-disc mosaics in this work to generate mock-SUIT images in the NB3 passband with a filter combination of CF1, which has a central wavelength of 2796~{\AA}. To construct mock SUIT images, we perform the following steps:
\begin{figure}    
    \centerline{\hspace*{0.015\textwidth}
             \includegraphics[width=0.515\textwidth,clip=]{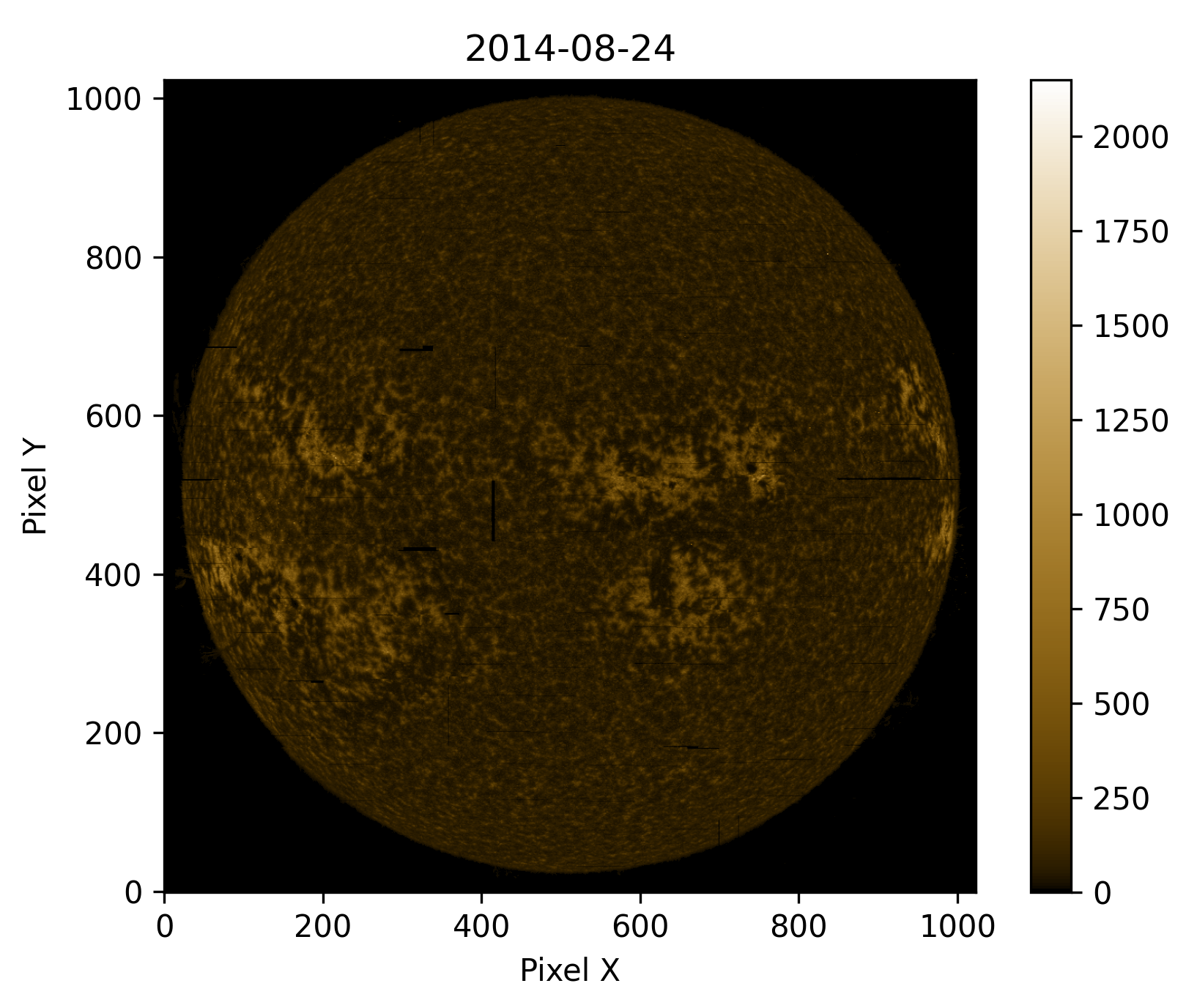}
             \hspace*{-0.03\textwidth}
             \includegraphics[width=0.515\textwidth,clip=]{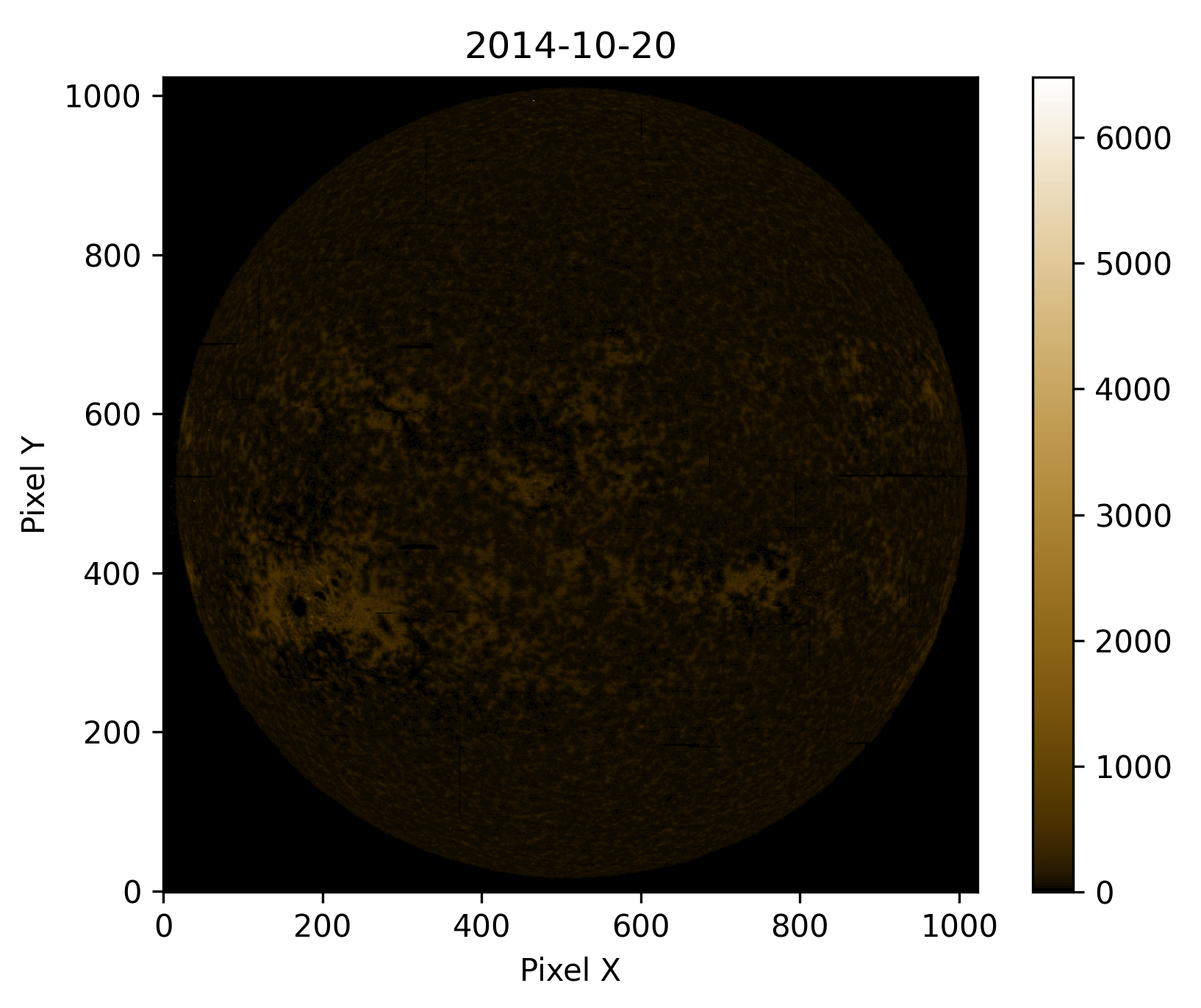}
            }
    \vspace{-0.37\textwidth}   
    \centerline{\Large \bf     
                \hspace{0.05 \textwidth}  \color{white}{(a)}
                \hfill}
    \vspace{0.3\textwidth}    
    \centerline{\hspace*{0.015\textwidth}
             \includegraphics[width=0.515\textwidth,clip=]{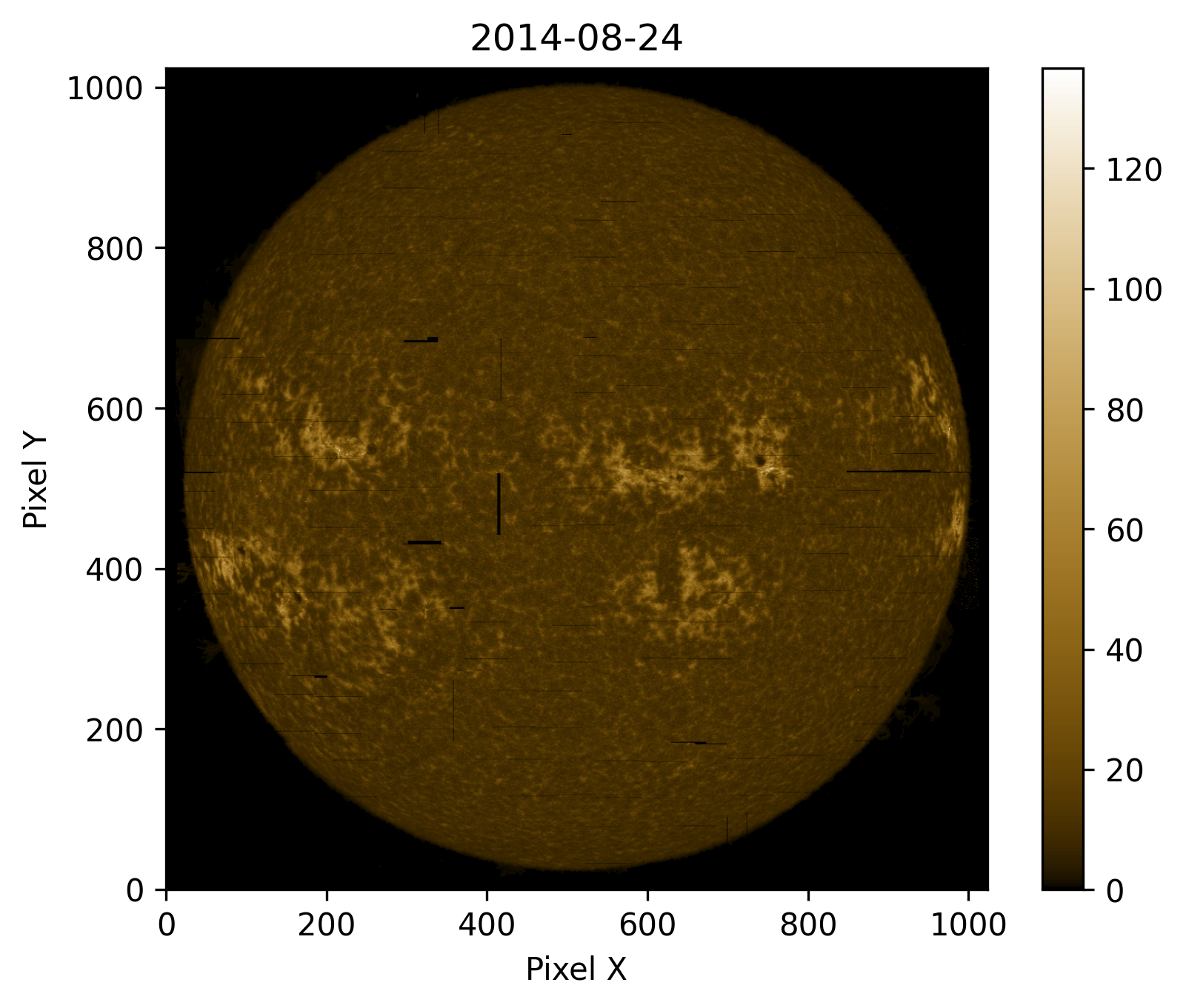}
             \hspace*{-0.03\textwidth}
             \includegraphics[width=0.515\textwidth,clip=]{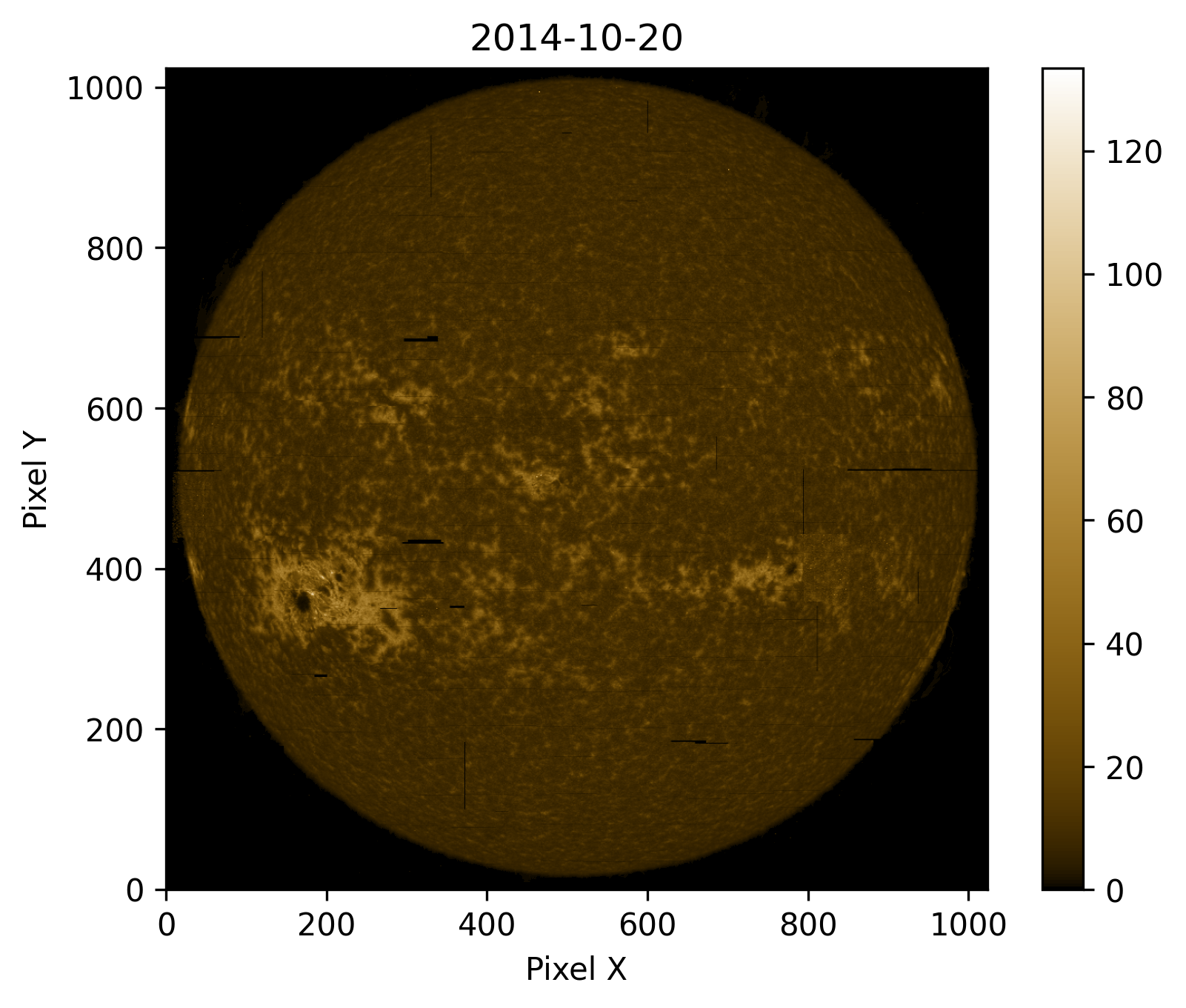}
            }
    \vspace{-0.37\textwidth}   
    \centerline{\Large \bf     
                \hspace{0.05 \textwidth} \color{white}{(b)}
                \hfill}
    \vspace{0.3\textwidth}    
    \centerline{\hspace*{0.015\textwidth}
             \includegraphics[width=0.515\textwidth,clip=]{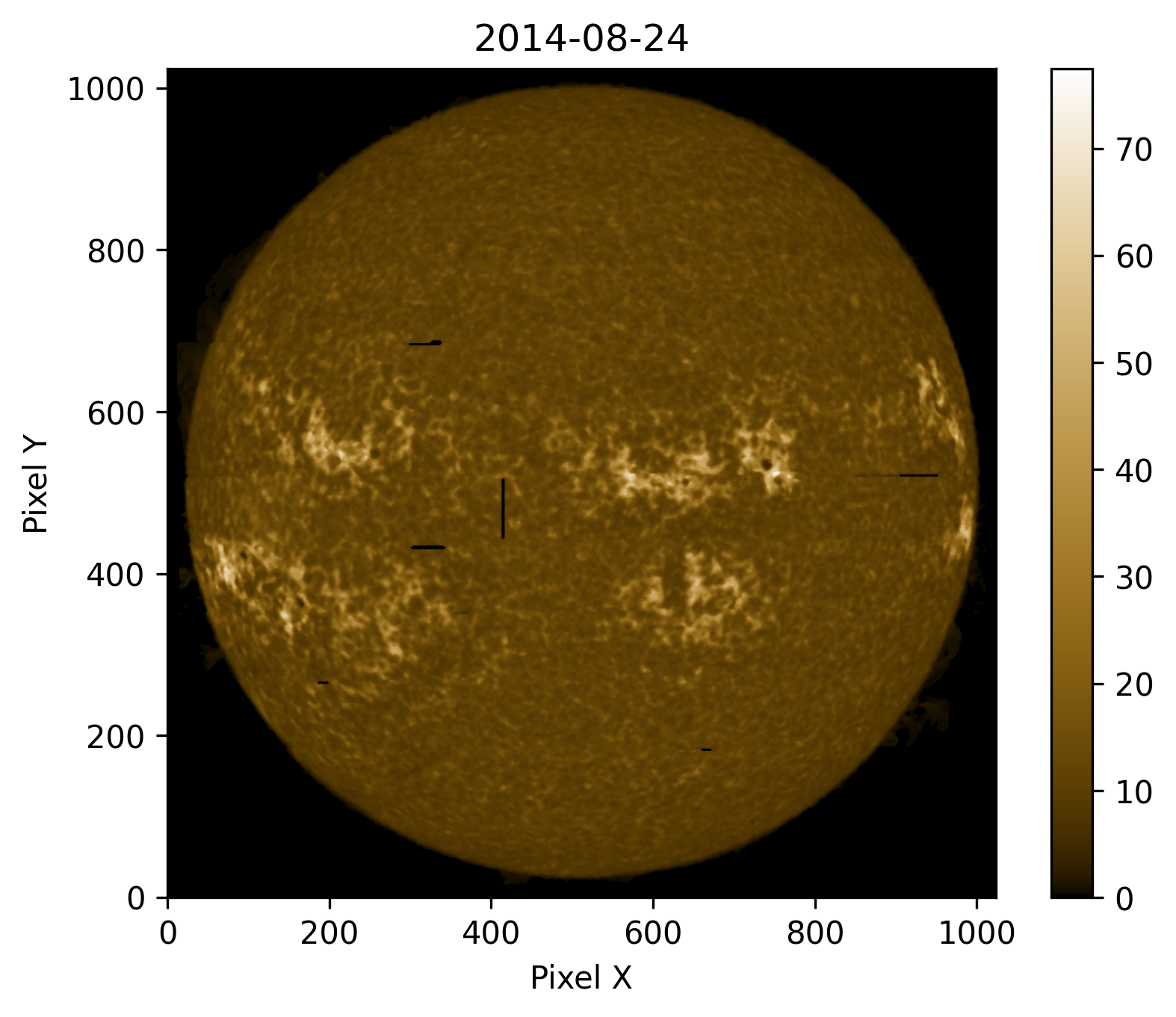}
             \hspace*{-0.03\textwidth}
             \includegraphics[width=0.515\textwidth,clip=]{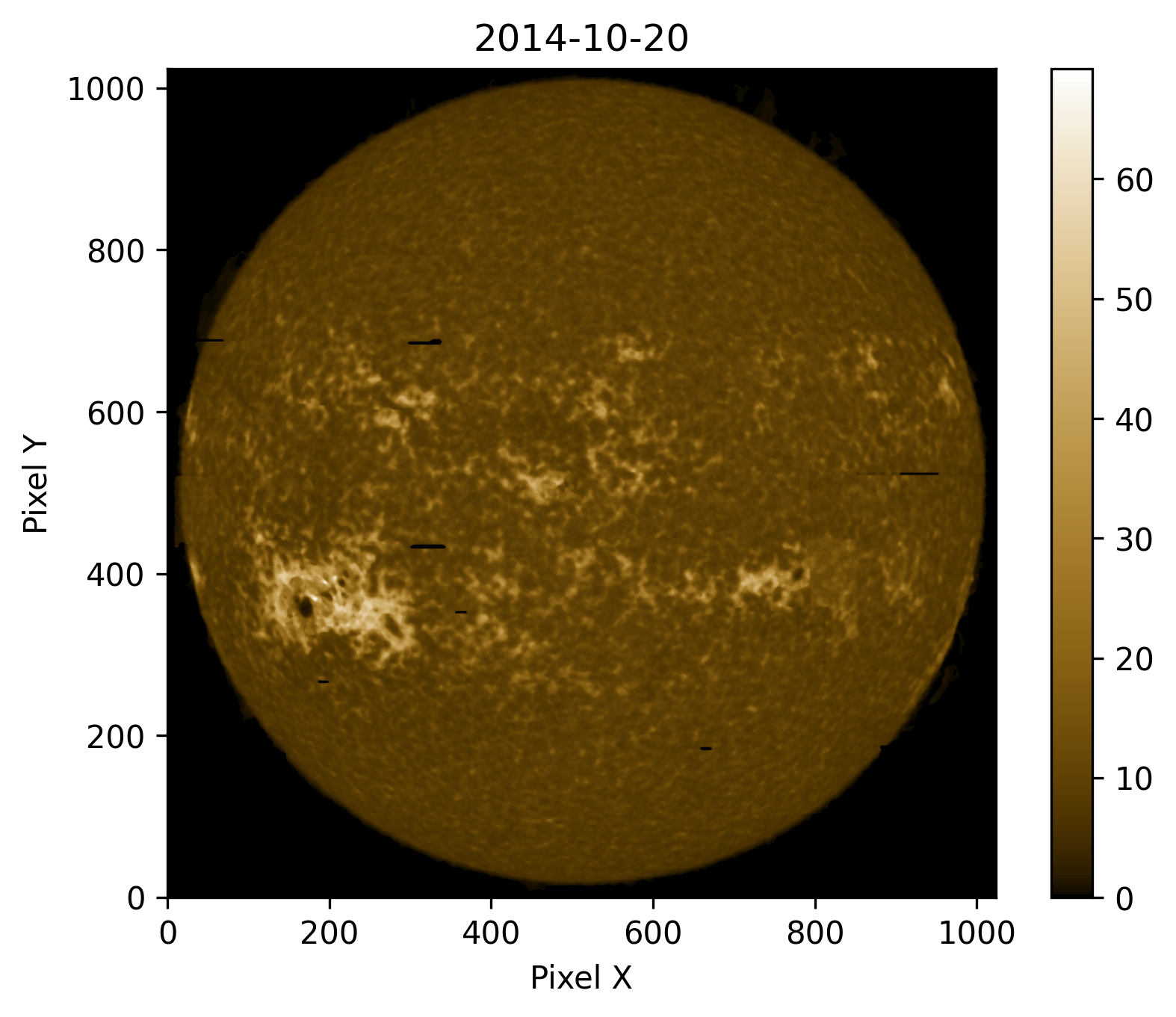}
            }
    \vspace{-0.37\textwidth}   
    \centerline{\Large \bf     
                \hspace{0.05 \textwidth} \color{white}{(c)}
                \hfill}
    \vspace{0.3\textwidth}    
    \caption{We present the preprocessing of IRIS mosaics to mock-SUIT data for two examples, presented in individual columns. From \textbf{\textit{Top to Bottom:}} Each row represents a step in the image processing pipeline.  \textbf{(\textit{a})} IRIS mosaic pixel values at the central wavelength of 2796~{\AA}. \textbf{(\textit{b})} The same IRIS mosaics folded with the normalized NB3 effective area (\textit{bottom row}).  \textbf{(\textit{c})} Image with cosmic ray artifacts removed using a 5$\times$5 median filter. }
    \label{fig:suitmock1}
\end{figure}

\begin{figure}
    \centerline{\hspace*{0.015\textwidth}
             \includegraphics[width=0.515\textwidth,clip=]{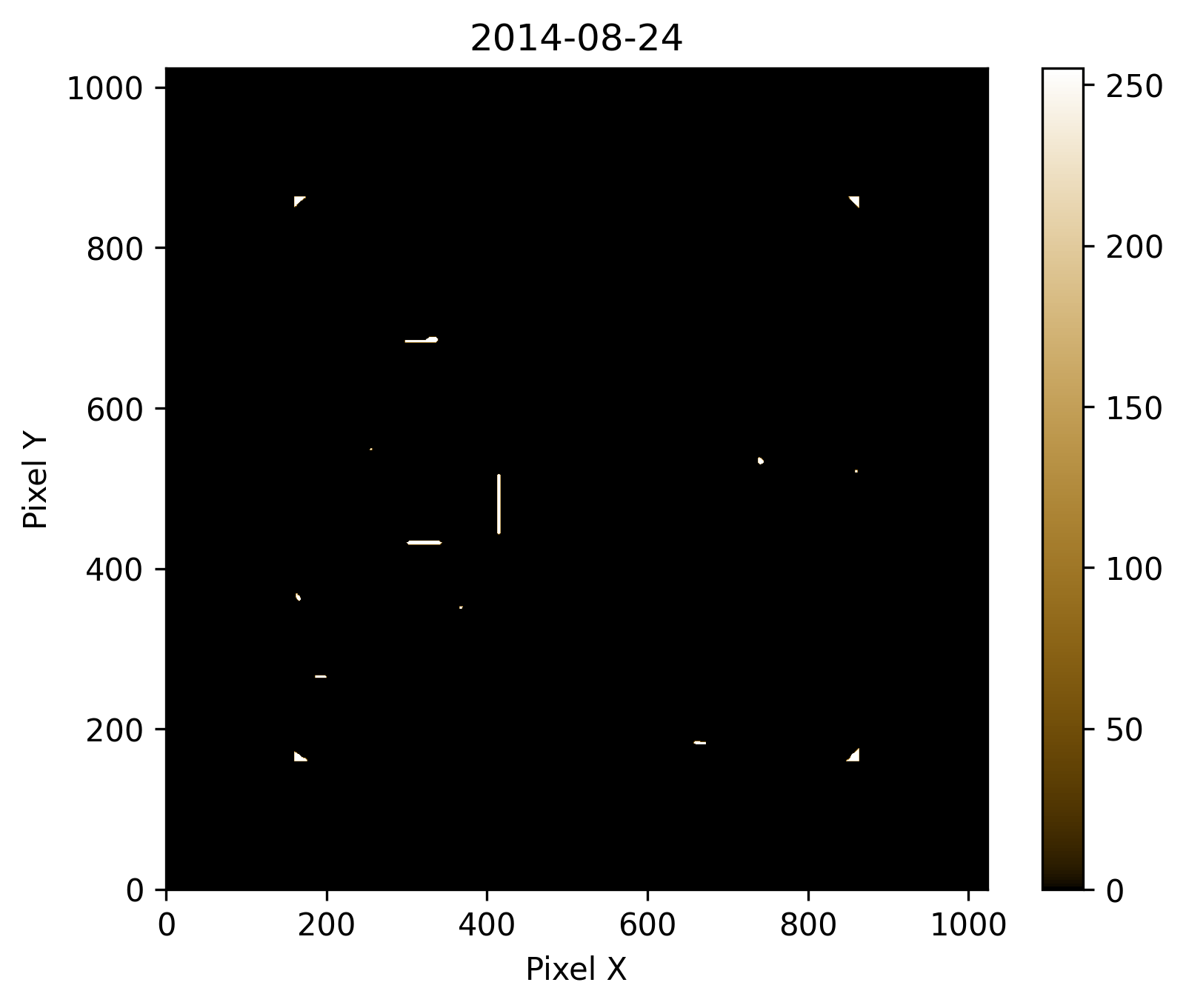}
             \hspace*{-0.03\textwidth}
             \includegraphics[width=0.515\textwidth,clip=]{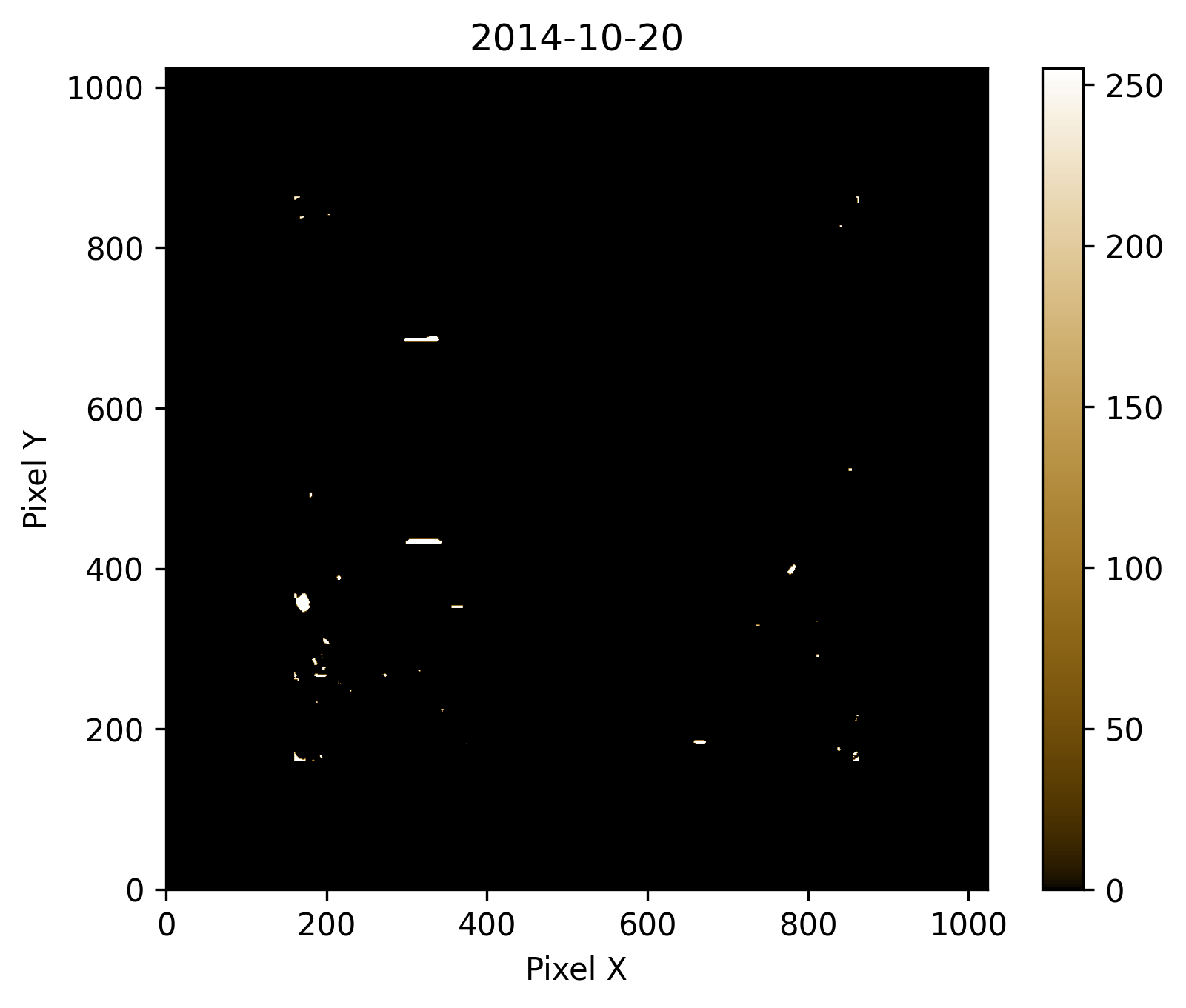}
            }
    \vspace{-0.37\textwidth}   
    \centerline{\Large \bf     
                \hspace{0.05 \textwidth} \color{white}{(a)}
                \hfill}
    \vspace{0.3\textwidth}    
    \centerline{\hspace*{0.015\textwidth}
             \includegraphics[width=0.515\textwidth,clip=]{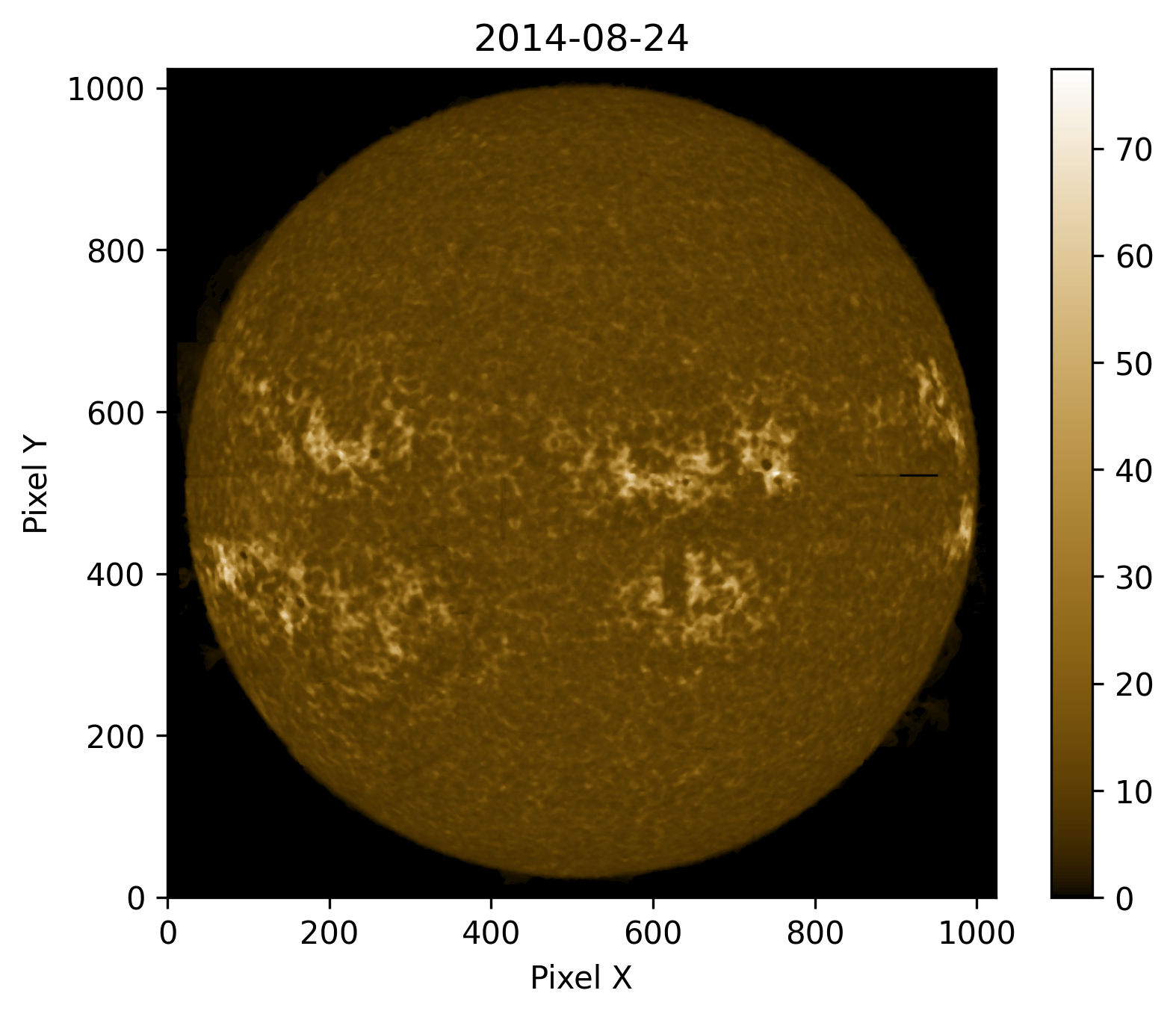}
             \hspace*{-0.03\textwidth}
             \includegraphics[width=0.515\textwidth,clip=]{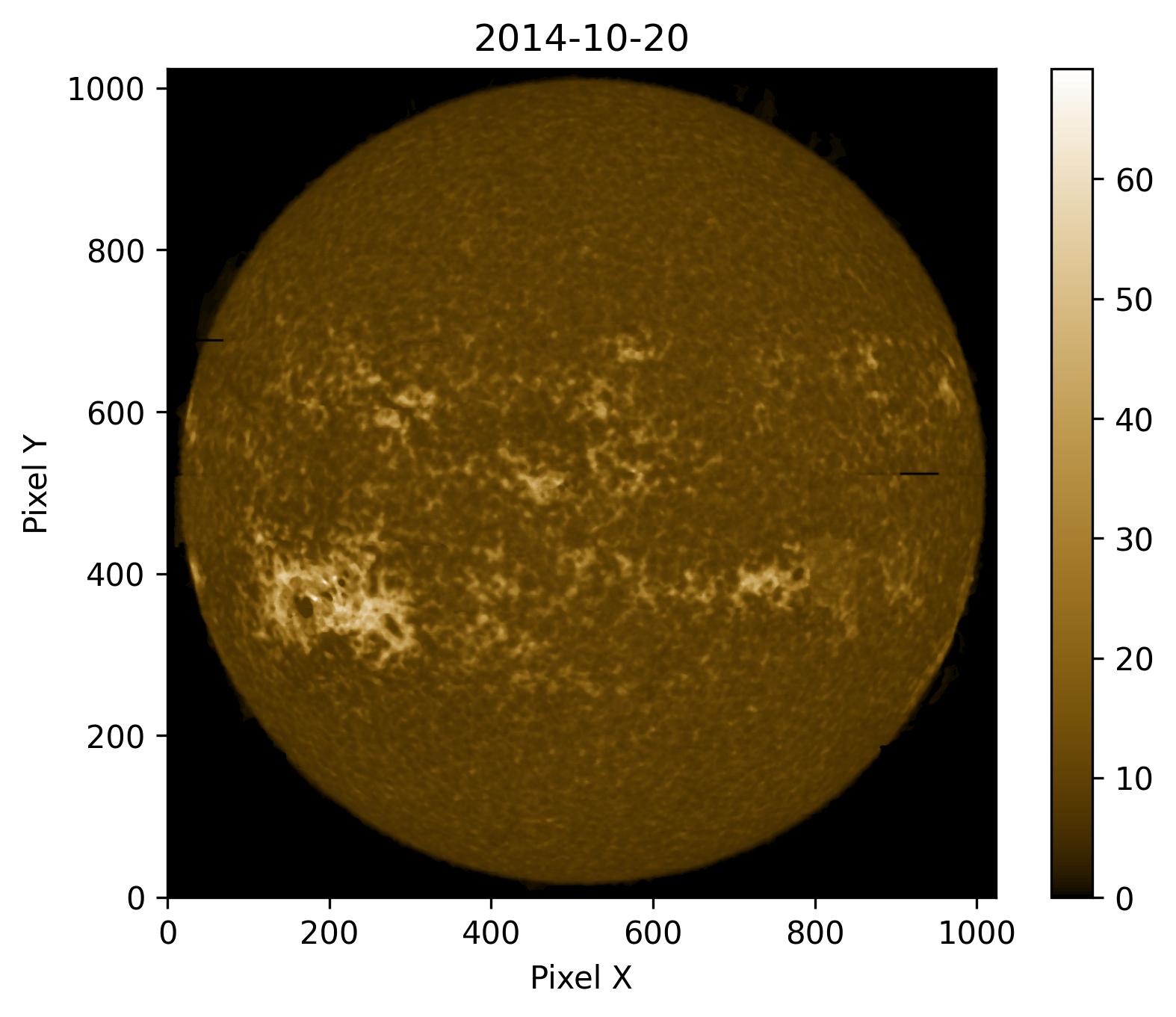}
            }
    \vspace{-0.37\textwidth}   
    \centerline{\Large \bf     
                \hspace{0.05 \textwidth}  \color{white}{(b)}
                \hfill}
    \vspace{0.3\textwidth}    
    \centerline{\hspace*{0.015\textwidth}
             \includegraphics[width=0.515\textwidth,clip=]{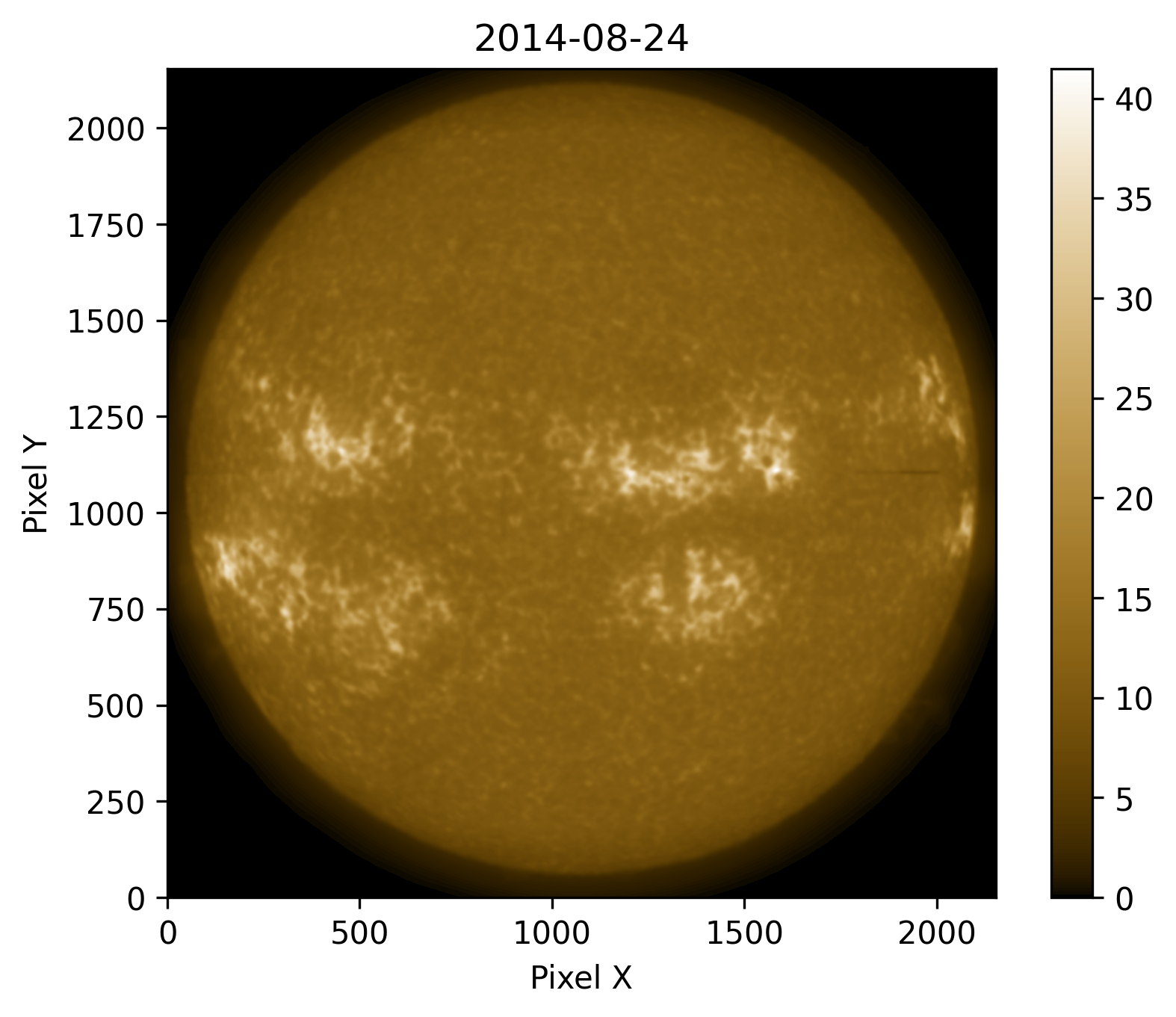}
             \hspace*{-0.03\textwidth}
             \includegraphics[width=0.515\textwidth,clip=]{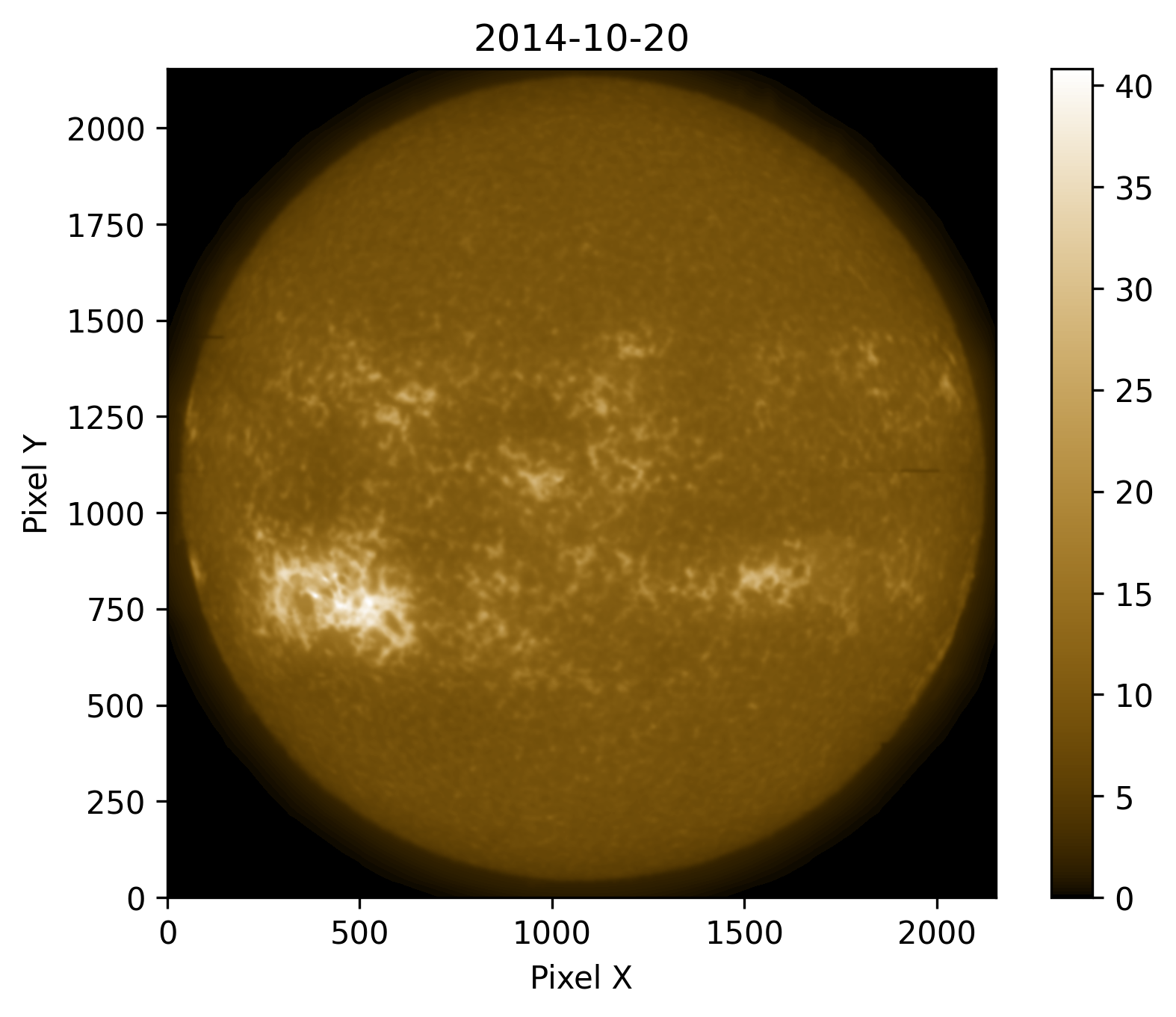}
            }
    \vspace{-0.37\textwidth}   
    \centerline{\Large \bf     
                \hspace{0.05 \textwidth} \color{white}{(c)}
                \hfill}
    \vspace{0.3\textwidth}    
    \centerline{\hspace*{0.015\textwidth}
             \includegraphics[width=0.515\textwidth,clip=]{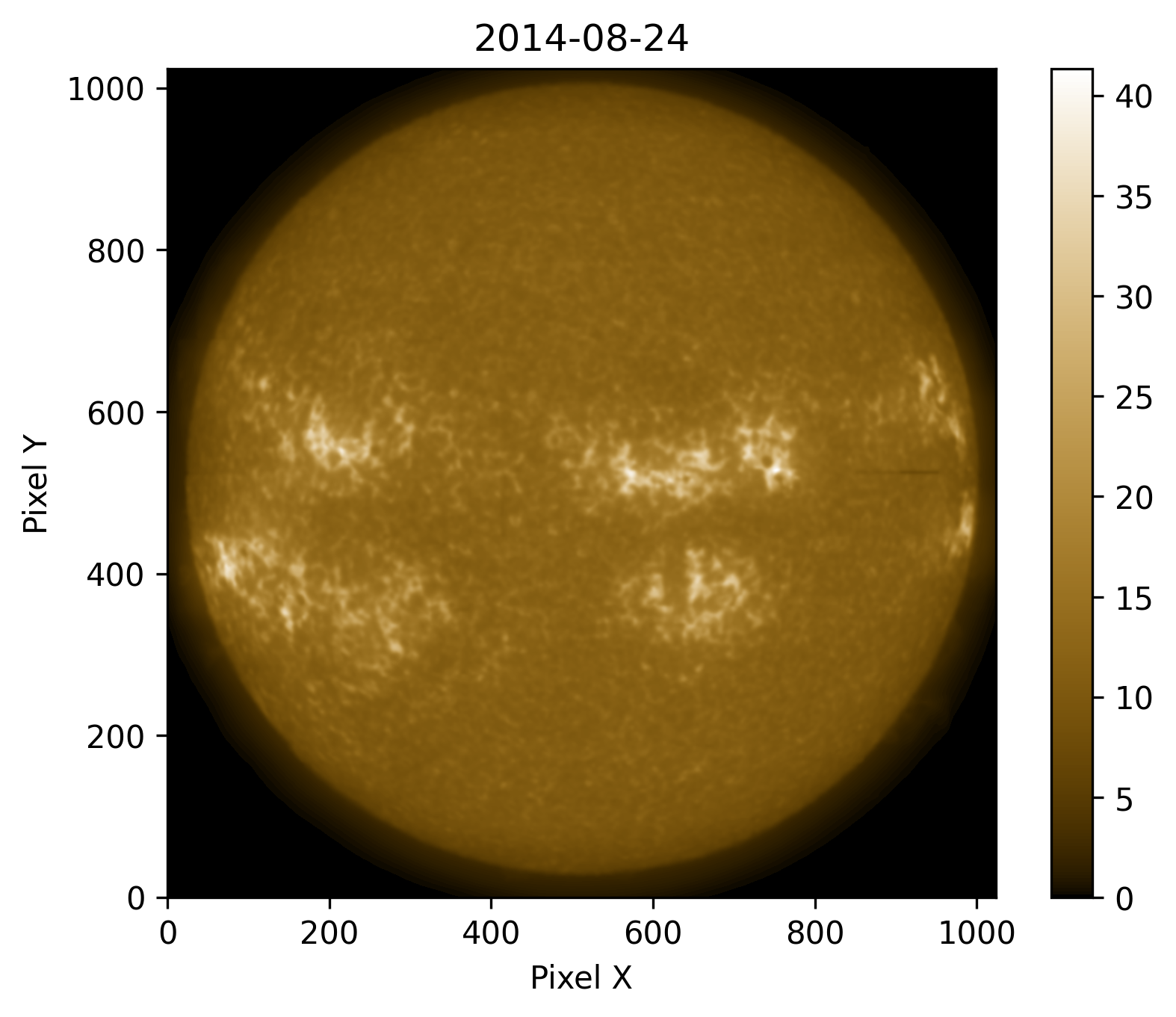}
             \hspace*{-0.03\textwidth}
             \includegraphics[width=0.515\textwidth,clip=]{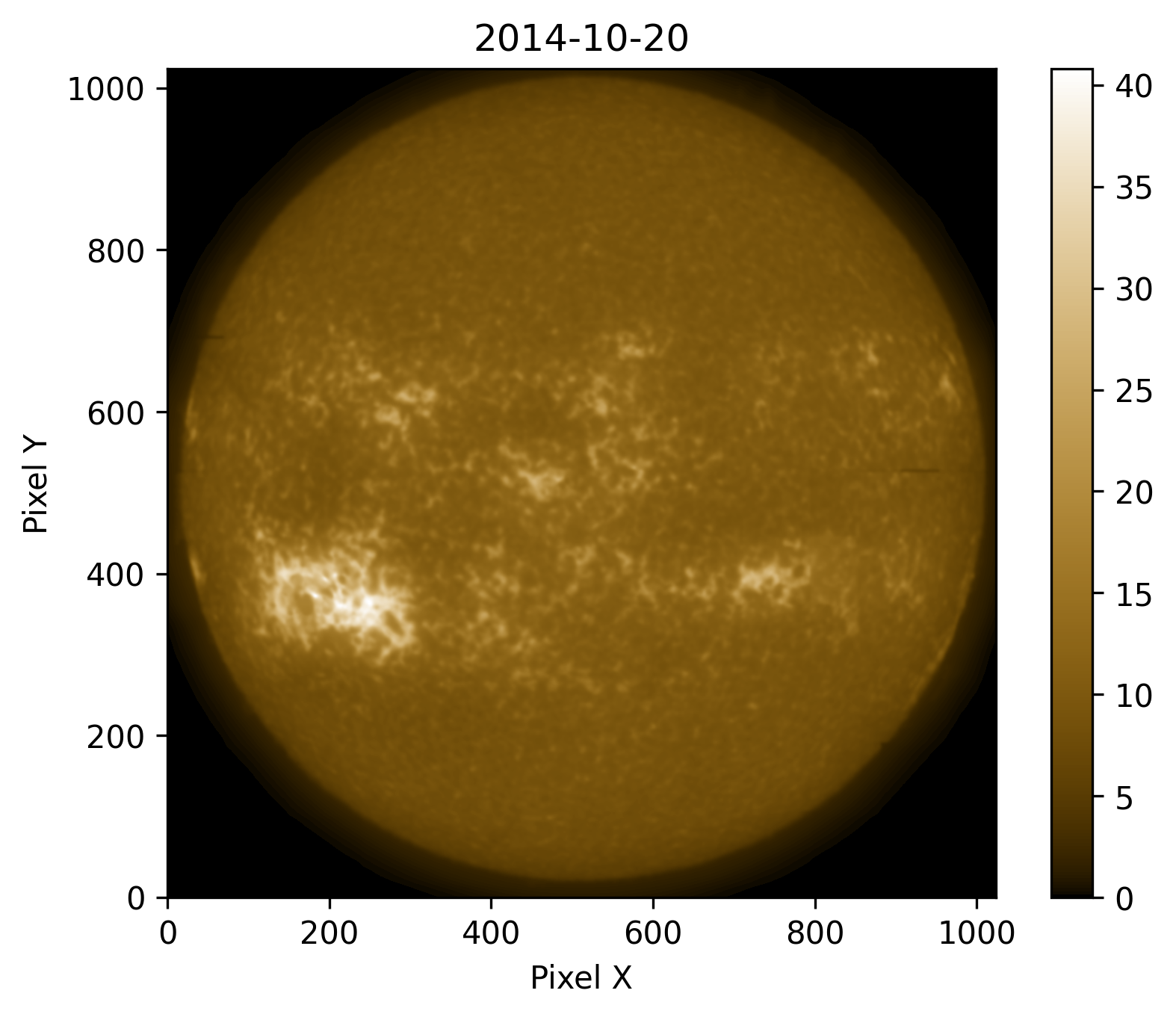}
            }
    \vspace{-0.37\textwidth}   
    \centerline{\Large \bf     
                \hspace{0.05 \textwidth} \color{white}{(d)}
                \hfill}
    \vspace{0.3\textwidth}    
    
    \caption{Continued from Figure~\ref{fig:suitmock1}. \textbf{\textit{Top to Bottom:}} Each row represents a step in the image processing pipeline.   \textbf{(\textit{a})} Binary mask generated through inverse thresholding. \textbf{(\textit{b})} The corresponding inpainted images. \textbf{(\textit{c})} The inpainted images after convolution with the SUIT NB3 PSF and rebinning to a resolution of 0.7'' per pixel. \textbf{(\textit{d})} The final mock SUIT images after further binning to a resolution of $1024\times1024$.}
    \label{fig:suitmock2}
\end{figure}

\begin{enumerate}
    \item Normalize the SUIT effective area to the maximum value. The normalized effective area, along with the wavelength range of IRIS observations is shown in Figure~\ref{fig:suit_effective_area}.
    
    \item This normalized effective area is multiplied with IRIS mosaic pixel values, and summed across all wavelength bins. Two examples of wavelength-summed IRIS Mg II k mosaics are shown in Figure~\ref{fig:suitmock1}b. These two images will serve as our test bed for demonstrating various cleaning operations and mock-SUIT image generation performed through this section.
    \item All IRIS mosaics are affected by spikes due to cosmic rays, which are not removed by default. To remove such cosmic ray spikes, we used a median filter \citep{median_filtering_algorithm} with a kernel size of 5 pixels.
    The median-smoothed images are shown in Figure~\ref{fig:suitmock1}c.
    \item The IRIS mosaics have dark horizontal or vertical lines due to mosaic stitching. To remove these artifacts, we performed a combination of Binary Inverse and Otsu thresholding~\citep{Otsu} to identify the dark pixels, which we treated as background. Binary Threshold tends to separate foreground and background with a single value, while Otsu's threshold algorithm seeks to find the intensity value that maximally separates intensity distribution into two Gaussians. The intensity masks are displayed in Figure~\ref{fig:suitmock2}a. We use these masks to perform inpainting~\citep{Inpainting} using the Navier Stokes function from OpenCV. At its core, this method the lines of constant intensity (i.e. Isophotes) constantly from the exterior into the area that has to be painted.  It considers the image intensity as a 2-D incompressible flow's ``stream function''. The Laplacian of the image intensity represents the fluid's vorticity, which is carried into the area that has to be painted by a vector field that is determined by the stream function. The resulting algorithm's goal is to match gradient vectors at the edge of the inpainting zone while maintaining isophotes. A detailed explanation of the method is provided in \citep{Inp2}. The data obtained after inpainting are shown in Figure~\ref{fig:suitmock2}b.We note that these techniques help reduce, and may not necessarily eliminate the raster scan joints, as evident from Figure~\ref{fig:suitmock2}a. However, since these effects are not labeled in the ground truth, we expect our model to learn and ignore them during the training process. 
    
    \item The angular resolution of IRIS mosaics is 0.33''/px along the $y$-axis and 1.99''/px along the $x$-axis. We interpolate the image to ensure that the angular resolution of both axes was equal to 0.33''/px. We then convolved the interpolated image with the measured Point Spread Function (PSF) values corresponding to SUIT NB3, as shown in Figure \ref{fig:suitmock2}c. We then degrade the images to 0.7''/px and bin them to 1024$\times$1024 spatial resolution to obtain the final image, as shown in Figure \ref{fig:suitmock2}d .  
    \end{enumerate}
\subsection{Data Processing}
\label{sec:data_preprocessing}
We manually label the prepared dataset using Roboflow~\footnote{\href{https://roboflow.com/}{roboflow.com}} by identifying the four features of interest: plages, filaments, sunspots, and off-limb structures.

The annotation is performed in the Oriented Bounding Box format (OBB)~\citep{yolov8_ultralytics}~\footnote{\href{https://docs.ultralytics.com/datasets/obb/}{docs.ultralytics.com/datasets/obb/}}. In this labeling format, each box is defined by the coordinates at the four vertices of the bounding box, and our model outputs the coordinates of the bounding box for each feature. 

\begin{figure}[!ht]
    \centering
    \includegraphics[width=\linewidth]{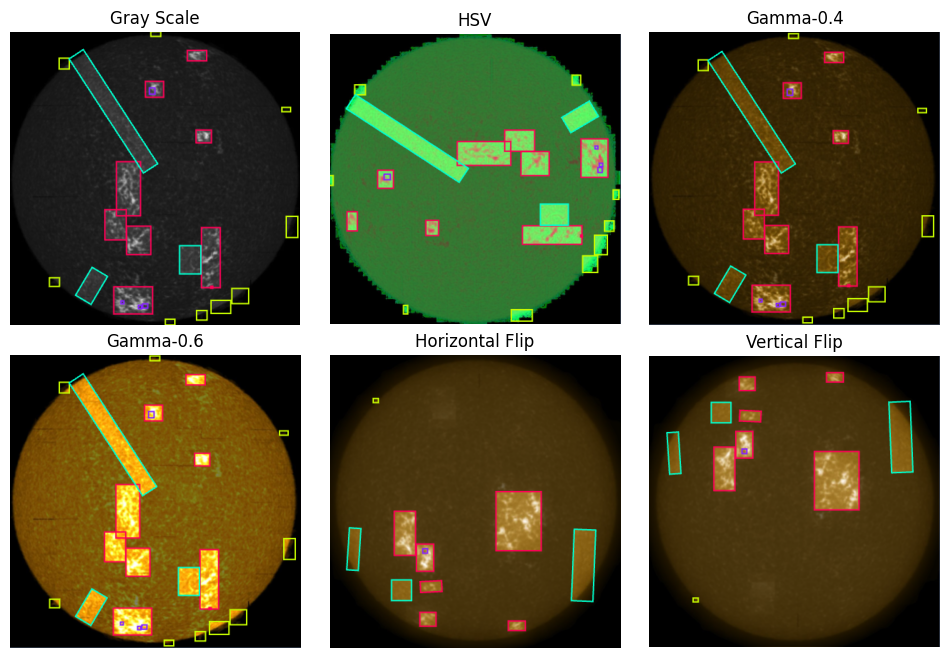}
    \caption{Example representative augmentation for an example mock SUIT image. Starting from the top-left and moving clockwise, the transformations include: Gray Scale conversion, HSV conversion, $\gamma = 0.4$, vertical and horizontal flip, and $\gamma = 0.6$.}
    \label{fig:augmentations}
\end{figure}

The original mock-SUIT images are $\approx 200$ images, which contain both psf-convolved and non psf-convolved images. While the usual norm is to consider the original data arrays, 200 images do not constitute a sufficient dataset for training a deep learning model. Hence, to perform effective data augmentation, we save the data as lossless JPG images, and perform several data augmentations as listed below. This results in $\approx3056$ images, which we use for training the model. The augmentation performed on the training images are:
\begin{itemize}
    \item Gray Scale Conversion: The color image is converted to a grayscale image.
    \item HSV conversion: HSV (Hue, Saturation, Value) conversion is a color representation model that separates image color information into three components: hue (the color type), saturation (the intensity or purity of the color), and value (the brightness of the color). We converted the images from  RGB to HSV space, where hue is mapped between 0 and 179, each channel having a unique value, while saturation varies between 0 and 255. 
    \item Gamma Variations: The brightness of the images are varied by changing the $\gamma$ value as per Eq. ~\ref{eqn:Gamma}. We varied $\gamma$ as 0.2 0.4, 0.6, 0.8 and chose the best looking ones. The augmentations were performed as: 
    \begin{equation}
        \centering
        \label{eqn:Gamma}
            image = [image]^\gamma
    \end{equation}
    These gamma variations serve to make the model robust to intensity variations occurring due to in-flight performance of SUIT. However, we do not select the full range of $\gamma$ for augmentation, and select only the images which do not look starkly different from the source images, through visual inspection.
    \item Vertical and Horizontal Flips.
\end{itemize}
The resultant set of augmentations are shown in Figure~\ref{fig:augmentations}.
\begin{table}[!ht]
    \begin{tabular}{lllp{6cm}p{6cm}}
    \hline
      Dataset &  No. of data samples  \\ \hline
      Pre-augmentation data & 217 \\
      Post augmentation training data & 3056 \\
      Post augmentation  validation (JPG) data & 297 \\
      Validation data (FITS) & 123\\
      Test data - SUIT Data (FITS) & 92 
       \\
      \hline	
    \end{tabular}
    \caption{Number of samples in the dataset.}
    \label{tab:dataset}
\end{table}

We evaluate the model on a validation set consisting of 123 samples, in FITS format. Note that to obtain a 3-channel image of our model, we replicate the mock SUIT image into all the three RGB channels. These data samples are used to evaluate the model weights. Finally, the testing set in our case are the actual SUIT Level 1 2k data. The details for the complete dataset are mentioned in Table \ref{tab:dataset}.

SUIT Level 1 data is science-ready FITS images, generated from Level 0 (uncorrected) data. Bias correction, flat field correction, scattered light removal, and dark removal are performed, followed by orientation correction and WCS implementation of the Sun images. \cite{testandcalib}.

Once the data is ready, we normalize the data across the mock-SUIT and SUIT data. We subtract each image by its minimum value, divide by the range, and then multiply it by 255 to ensure compliance with 8-bit image representation. 
\section{\textbf{Model Architecture, Selection and Evaluation}}
\label{sec:MAE}
{\spc} is based on the architecture of You Only Look Once (YOLO) \citep{v8}\citep{v9}, containing a backbone of a Convolutional Neural Network (CNN). The YOLO series of models contains a backbone, neck, and head. It ingests RGB images and outputs the bounding boxes for each feature, along with the probability of the feature in consideration. Here, we have utilized two versions of YOLO, v8~\citep{v8,yolov8_ultralytics} and v9~\citep{v9}. In the YOLO-v8, we also consider the v8-mobb and v8-m models. The v8-mobb model considers the input coordinates in the OBB format. In contrast, the v8-m model considers the bounding boxes in traditional axis-aligned scheme. We refer to the v8 papers for details on the model architectures~\citep{v8,yolov8_ultralytics}.YOLOv8-mobb is  designed to detect objects with arbitrary orientations by predicting rotated bounding boxes instead of traditional axis-aligned ones.

Several state of the art architectures like YOLO, Faster R-CNN, EfficientDet, and Single Shot Detector (SSD) exist to perform the task at hand. However, our requirements were defined as enabling computationally cheap, near real-time detections of features spanning various length scales. Faster R-CNN, while highly accurate, could be sometimes be computationally expensive due to its two-stage detection process \citep{FRCNN}. EfficientDet provides a trade-off between speed and accuracy, but its real-time efficiency depends on the specific variant used \citep{EDET}. Single Shot Detector (SSD) is an efficient single-stage detection alternative but struggles with accuracy, particularly for small-scale features \citep{SSD} like in our case, compared to YOLO. Hence, we went ahead with a YOLO for this work.

We consider the following target classes: ``Plages", ``Sunspots", ``Filaments" and ``Off-limb features". We trained the models for 500 epochs, with a batch size of 8, and a combination of multiple loss functions. YOLOv8 has a loss function consisting of three terms: bounding box regression loss (${L}_{\rm box}$), confidence loss (${L}_{\rm conf}$), and classification loss (${L}_{\rm class}$) as shown in Eq.~\ref{eqn:v8loss}.
\begin{equation}
    \mathcal{L}_{\text{YOLOv8}} = \mathcal{L}_{\rm box}+ \mathcal{L}_{\rm conf} + \mathcal{L}_{\rm class}
    \label{eqn:v8loss}
\end{equation}

In the v8-obb model, the loss function is defined as: $$\mathcal{L}_\text{OBB} = \mathcal{L}_{\text{box}}+\mathcal{L}_{\rm angle}(\theta_i, \hat{\theta}_i)  + \mathcal{L}_{\rm conf} + \mathcal{L}_{\rm class},$$ where the loss term includes an additional term to account for the orientation of the bounding boxes. The additional loss term $\mathcal{L}_{\rm angle}(\theta_i, \hat{\theta}_i) \Big)$ is defined as $$\mathcal{L}_{\text{\rm angle}}(\theta_i, \hat{\theta}_i) = \min \big(|\theta_i - \hat{\theta}_i|, 2\pi - |\theta_i - \hat{\theta}_i|\big),$$ where the loss term compares the angular error between the predicted and ground truth angles (\(\hat{\theta}_i\) and \(\theta_i\)). This angular loss is designed to handle the periodic nature of angles, ensuring smooth error 
 calculations even at boundary conditions (e.g., \(0^\circ\) vs \(360^\circ\)).

The key improvements in YOLOv9, when compared to v8, include Programmable Gradient Information (PGI), which helps preserve critical information across layers to improve learning, and the Generalized Efficient Layer Aggregation Network (GELAN).YOLOv9's loss function is a  combination of CIoU loss by default as mentioned above, objectness loss and classification loss. \\
The final total loss function for YOLOv9 is a weighted combination of these components:

\[
\mathcal{L}_{\text{YOLOv9}} = \mathcal{L}_{\text{\rm CIoU}} + \lambda_{\text{\rm obj}} \mathcal{L}_{\text{\rm obj}} + \lambda_{\text{\rm class}} \mathcal{L}_{\text{\rm class}}
\]
\\
where \(\lambda_{\text{\rm obj}}\) and \(\lambda_{\text{\rm class}}\) are hyperparameters that control the importance of the objectness and classification losses in the total loss.
\\
We used an ``auto'' setting for the optimizer in both YOLOv8~\citep{v8} and YOLOv9~\citep{v9}, which internally selects the optimizer based on the model architecture and training configuration. The initial learning rate here was set to 0.01.
\subsection{Model Evaluation}
\label{subsec:model_evaluation}
We use the following metrics to evaluate our model performance for various IoU:
\begin{itemize}
    \item Precision: A measure of the fraction of predicted positive instances that are actually positive, defined as:
            \begin{equation}
            \label{eqn:precision}
                \mathrm{Precision} = \frac{\mathrm{True Positive}}  {\mathrm{True Positive + False Positive}}.
            \end{equation}
    \item Recall: A measure of the fraction of actual positive instances that are correctly identified, defined as:
        \begin{equation}
        \label{eqn:recall}
        \mathrm{Recall} = \frac{\mathrm{True Positive}}{\mathrm{True Positive + False Negative}}.
        \end{equation}
    \item Mean Average Precision (MAP): AP is the average precision of each class, while MAP is the mean of AP scores across all classes, defined as:
        \begin{equation}
        \label{eqn:mAP}
        {
            \mathrm{MAP} = \frac{1}{N} \sum_{i=1}^{N} \mathrm{AP}_i.
        }        \end{equation}
    
\end{itemize}
\section{Evaluation Strategy: Statistical and Tamura Measures}
\label{sec:STF}
In this work, we have manually identified the bounding boxes for each feature in the mock SUIT images. For the given bounding boxes, we would like to have a more objective set of measures for determining the different regions of interest. Specifically, this is also needed in the absence of a GT in the SUIT images, to verify if the features we recover statistically are consistent with the class labels. Visually, these different morphological regions appear different because of differences in the intensity distribution. Hence, we compute statistics of intensity distribution within the bounding boxes. We define 10 such parameters, collectively calling them \textit{Feature properties}:
\begin{enumerate}
    \item Entropy: For a probability distribution $P(x)$, the Shannon entropy \citep{ent} is defined as: - $\sum P(x_i)\log P(x_i)$. This provides a measure of `non-peaked-ness' of the intensity distribution.
    \item Standard Deviation ($\sigma$):  $\sqrt{\frac{\sum (x_i - \bar{x})^2}{N}}$, where $\bar{x}$ is the mean intensity in the bounding box and N denotes the sample size.
    \item Skewness: It is a measure of the asymmetry of the pixel intensity distribution, and is the third central moment of an image. The skewness is calculated as:
    $$N \cdot \sum_{i=1}^{N} (x_i - \bar{x})^3/((N-1)(N-2) \cdot \sigma^3),$$
    where N, $x_i$, $\bar{x}$, $\sigma$ denotes the sample size, individual data points, mean, and standard deviation, respectively.
    \item Kurtosis: It is a measure of the occurrence rate and magnitude of the anomalies and is the fourth central moment. The kurtosis is measured as: $$N \cdot (N+1) \cdot \sum_{i=1}^{N} (x_i - \bar{x})^4 {(N-1) \cdot (N-2) \cdot (N-3) \cdot \sigma^4} - \frac{3 \cdot (N-1)^2}{(N-2) \cdot (N-3)},$$
    where N, $x_i$, $\bar{x}$, $\sigma$ denotes measure sample size, individual data points, mean, and standard deviation, respectively.

\end{enumerate}

Our other measures include the Tamura properties computed for each feature. The Tamura properties refer to measurements of differences in texture characteristics such as coarseness, contrast, roughness, etc.  In this work, we use the Tamura properties to identify textural patterns along 0, 45, and 90-degree orientations. We calculate the Gray Level Co-Occurrence Matrix (GLCM) \citep{GLCM} for each bounding box, and define our metrics. For each element $Pij$ in GLCM we define:
\begin{enumerate}
    \setcounter{enumi}{4}
    \item Contrast: $\sum P_{ij}(i-j)^\mathbf{2}$
    \item Homogeneity: $\sum P_{ij}/\mathbf{1}+P_{ij}(i-j)^\mathbf{2}$
    \item Dissimilarity: $\sum P_{ij}(i-j)$
    \item Energy: $\sqrt{\sum P_{ij}^\mathbf{2}}$
    \item Correlation: $\sum P_{ij}[(i-\mu i)(j - \mu j)] / \sqrt{\sigma i^\mathbf{2} \sigma j^\mathbf{2}}$
\end{enumerate}

These set of metrics are computed for each bounding box in the training, validation and testing sets, for interrogation of both the labels and model detections.

\section{Results}
\label{sec:results}
In this section, we present the results of our developed models, and also present the comparative study of \textit{Feature properties} for the Ground truth, Predictions, and SUIT data.

\subsection{Visual Detection} 
\label{subsec:VDS}
\subsubsection{Model Performance on Mock SUIT Data}
\label{subsec:STS}
We validate the final models YOLO-v8, YOLO v8-obb, and YOLO-v9 using the mock SUIT data, as mentioned in Section \ref{sec:data}. We present the validation on the JPG images. These numbers are presented in Table~\ref{tab:val_res_jpg}. From these metrics, we find the model v8-obb to be the best model for the task at hand. Hence, we consider the obb model as our main model for \spc. 

We now present performance of the v8-obb model on the FITS data, in Table \ref{tab:val_res_fits}, for each class label. The first row of Table~\ref{tab:val_res_fits} contains the summary statistic of all features, and can be compared with the first row of  Table~\ref{tab:val_res_jpg}. We find that the model performance to be better with JPG over FITS data. This arises because the JPG data is (i). Lossy, and (ii). Quantized as integers. Hence, very subtle variations in intensity are washed out due to the JPG compression, which results in a better performance in JPG over FITS data. 

\begin{figure}[!ht]
    \includegraphics[width=\linewidth]
    {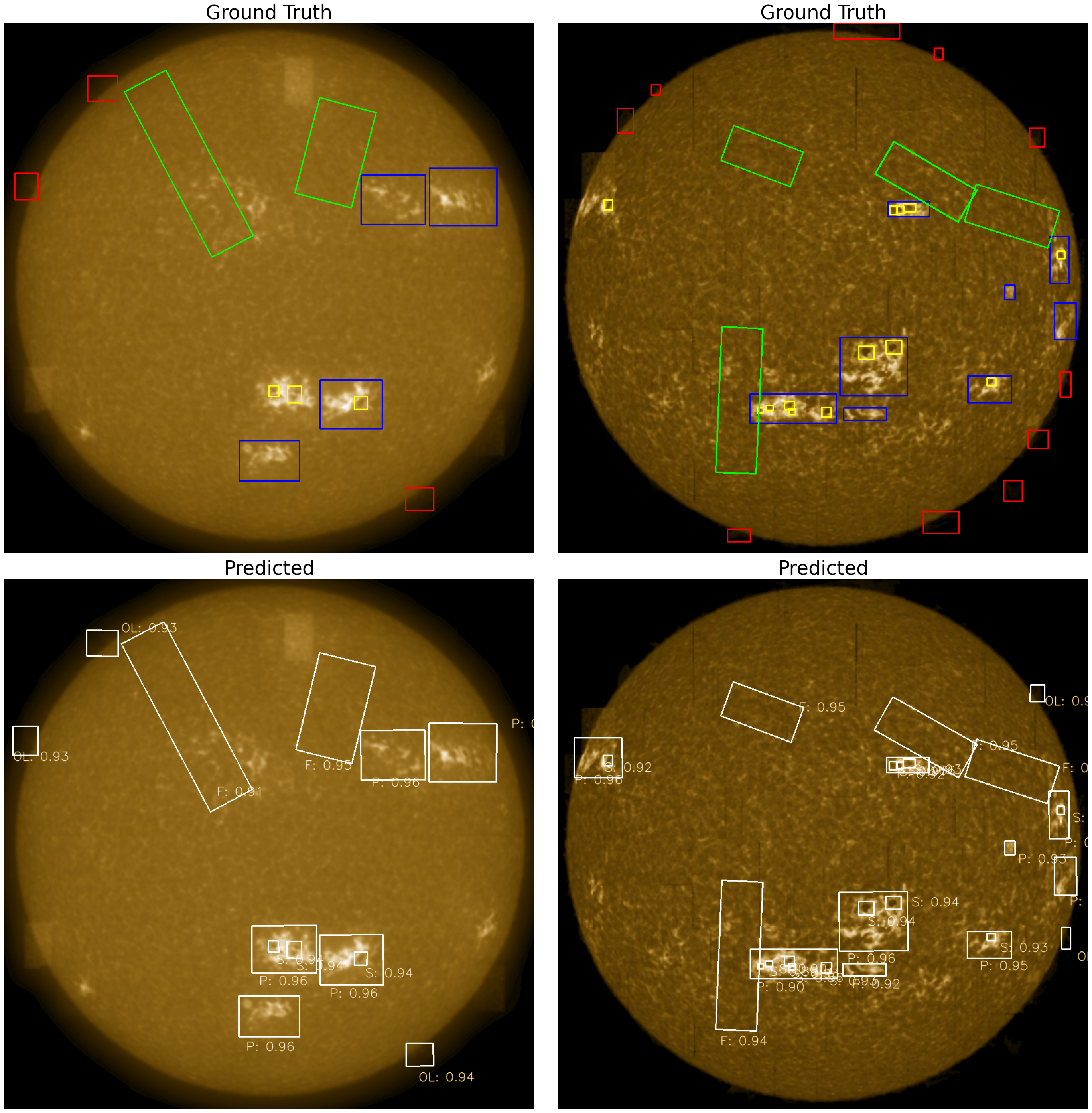}
    \caption{Predictions on SUIT MOCK FITS. F: Filaments, OL: Off Limb features, S: Sunspots P: Plages. The numbers in each box present the confidence score calculated for each feature between 0 and 1. The images shown were observed on 13 October 2013 \textit{(right column)} and 9 September 2021 \textit{(left column)}. The top panels
    show the manually labeled bounding boxes: F \textit{(green boxes)}, OL \textit{(red boxes)},
    S \textit{(yellow boxes)}, and P \textit{(blue boxes)}.}
    \label{fig:Inference_1}
\end{figure}

Comparing the model performance for different features in Table~\ref{tab:val_res_fits}, we find that the model shows excellent recall in most cases, except for off-limb regions. The precision is best for filaments and plages, and lesser for sunspots and off-limb regions. We present inference on two SUIT mock examples with the inputs taken in the native FITS format, and show them in Figure~\ref{fig:Inference_1}. The top panels show the manually labeled bounding boxes for four distinct solar features—Filaments (F, green boxes), Off-limb features (OL, red boxes), Sunspots (S, yellow boxes), and Plages (P, blue boxes). These labels were generated through the manual annotation process described in Section \ref{sec:data_preprocessing} and serve as the ground truth against which the model’s performance is compared. In the bottom panels, the YOLO v8-obb model’s predicted bounding boxes are overlaid on the same FITS images. As noted in Table~\ref{tab:val_res_fits}, off-limb regions are not detected reliably (low recall), while filaments and plages show higher precision than sunspots and off-limb features.

\begin{table}[!ht]
    \centering
    \begin{tabular}{lllp{3cm}p{3cm}}
    \hline
    & & & JPG  & \\
    \hline
      Models &  Precision  & Recall & Mean Average Precision [0-50] (MAP) & Mean Average Precision[ 50-95] (MAP) \\ \hline
      v8obb & 0.85077 &
            0.99546 & 0.97217 & 0.90361\\
      v8 & 0.78675 &0.98475 &0.94937 &0.84687 \\
      v9 & 0.79491 &0.97812 &0.94705&0.82481\\
      \hline	
      
    \end{tabular}
    \caption{\textbf{Comparison of the validation results on the mock SUIT JPG data for all models.}}
    \label{tab:val_res_jpg}
\end{table}
\begin{table}[!ht]
    \centering
    \begin{tabular}{lllp{3cm}p{3cm}}
    \hline
      Class labels &  Precision  & Recall & Mean Average Precision [0-50] (MAP) & Mean Average Precision[ 50-95] (MAP) \\ \hline
       all    &         0.788     & 0.863    &  0.874     & 0.819 \\
       filament   &      0.78 &      0.96  &    0.934 &     0.906 \\
       off-limb  & 0.875 &     0.542  &    0.751   &    0.63 \\
       plages  & 0.716 &     0.985 &      0.922  &    0.905 \\
       sunspot &  0.783 &  0.965 &     0.889  &   0.834 \\
      \hline	
    \end{tabular}
    \caption{\textbf{Validation results for the selected v8-obb model on the mock SUIT FITS data.}}
    \label{tab:val_res_fits}
\end{table}

\subsubsection{Model Deployed on the Real SUIT Data}

\begin{figure}[!ht]
    \centering
    \includegraphics[width=0.45\linewidth]{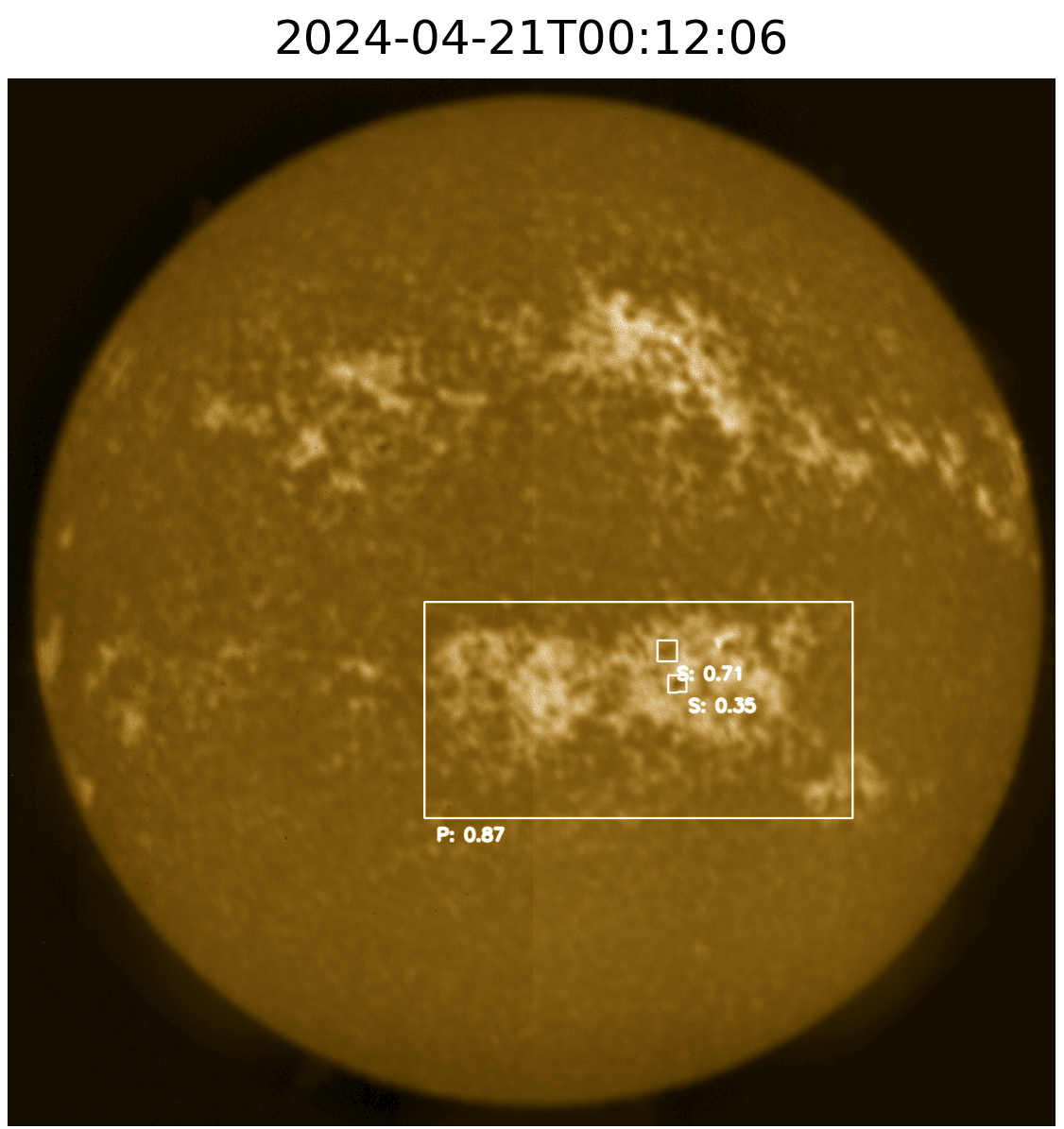}
     \includegraphics[width=0.45\linewidth]{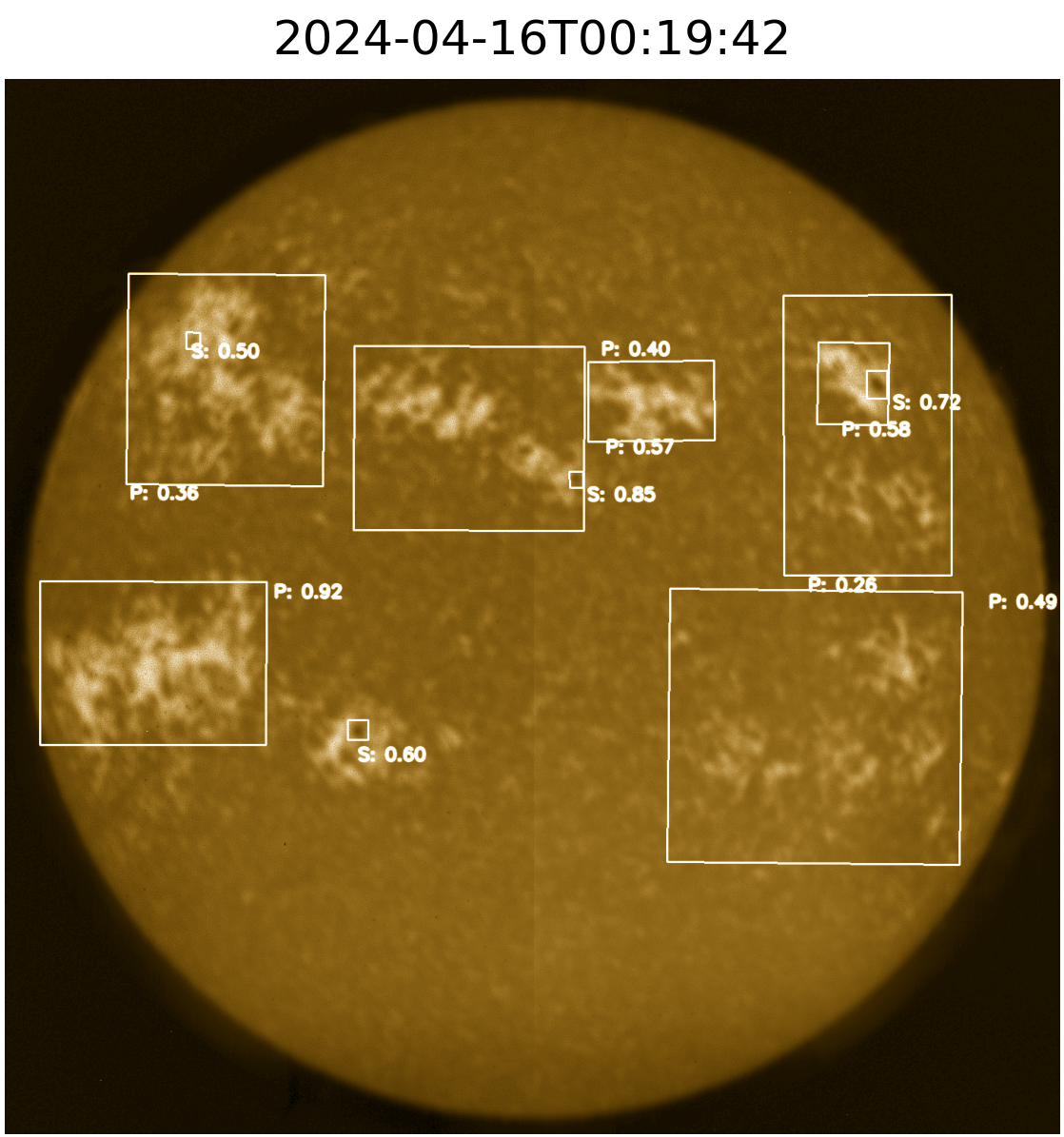}
     \includegraphics[width=0.45\linewidth]{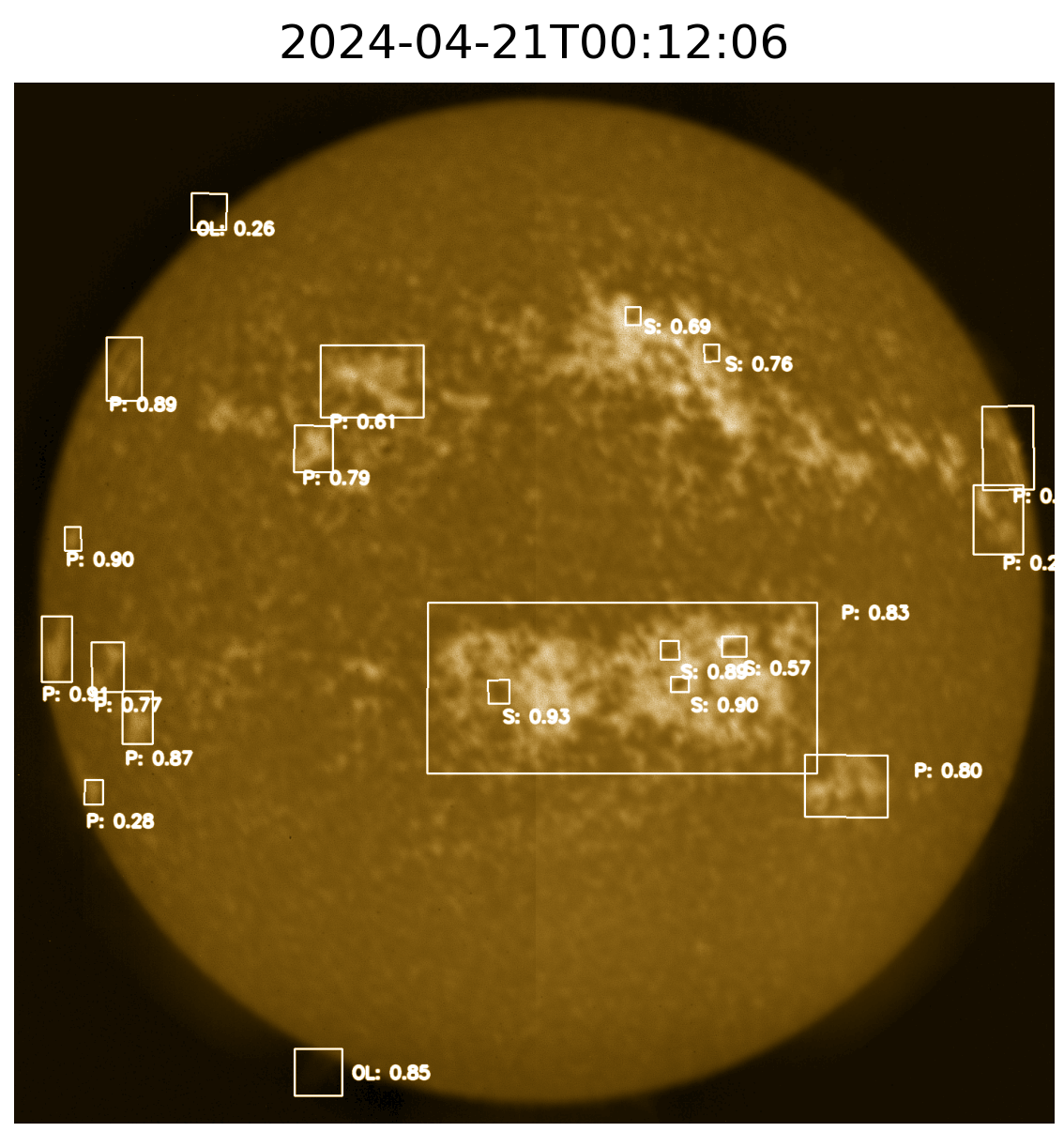}
     \includegraphics[width=0.45\linewidth]{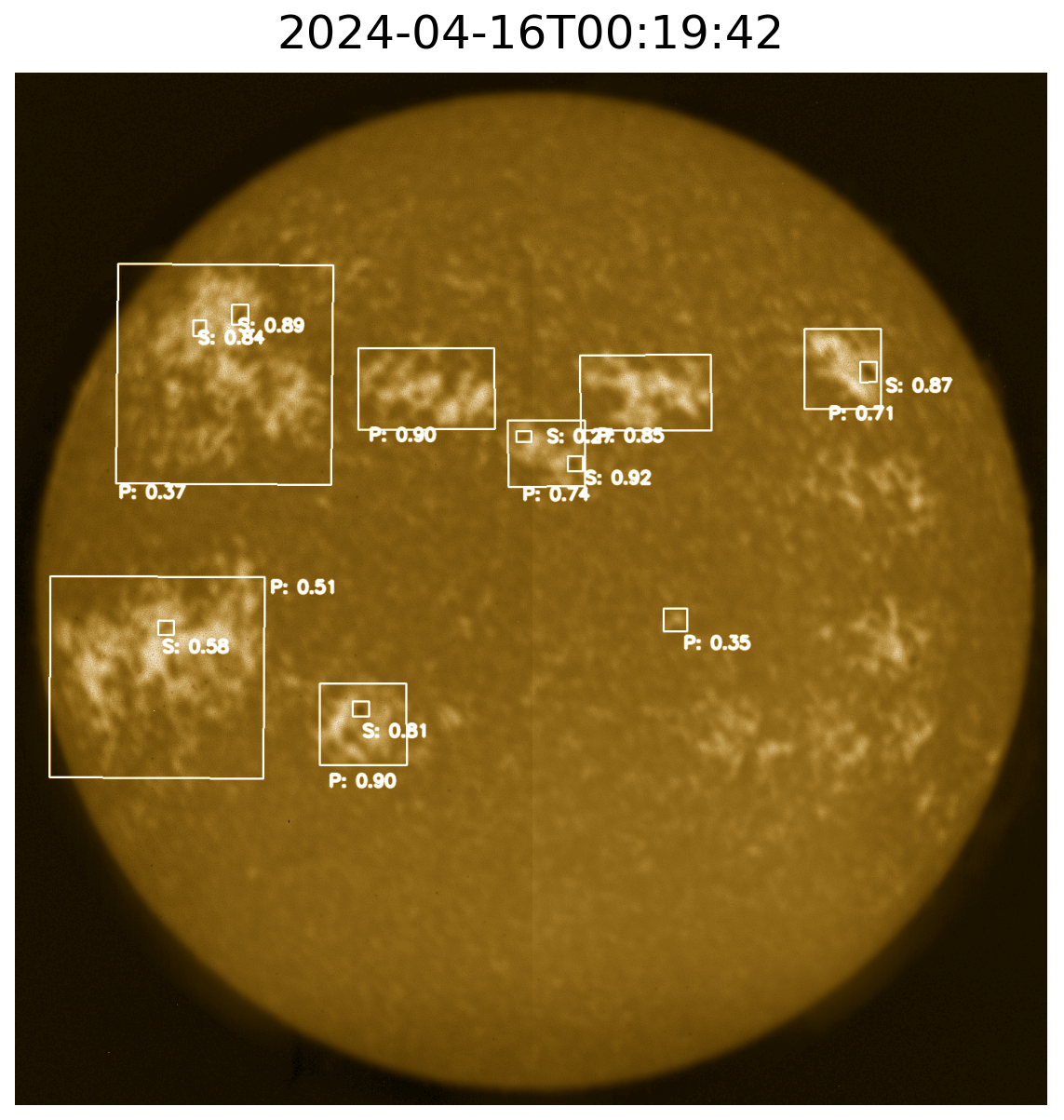}
    \includegraphics[width=0.45\linewidth]{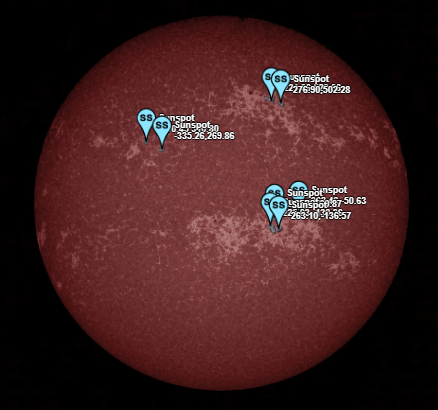}
    \includegraphics[width=0.443\linewidth]{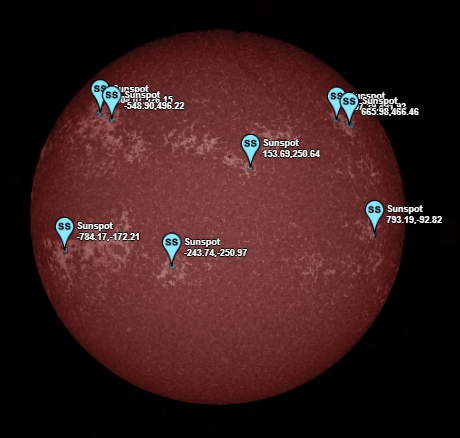}

    \caption{Inference on SUIT Images \textit{(top two panels)} and reference AIA 1700 images\textit{ (bottom panel)}. The features are given as: F: Filaments, OL: Off Limb features, S: Sunspots P: Plages; Confidence Score calculated for each event, whose value lies between 0 and 1. The \textit{top row }corresponds to inference using the min-max image rescaling, while the \textit{middle row} corresponds to inference using im/8 rescaling.}
    \label{fig:SUIT_DATA}
\end{figure}

In Figure~\ref{fig:SUIT_DATA}, we present the performance of our best model which is v8m-obb mentioned above on the observed SUIT data. The panel contains two sets of images compared with sunspot identification from the 1700~{\AA} passband of the Atmospheric Imaging Assembly (AIA)~\citep{Lemen2012,BoeEL_2012} onboard Solar Dynamics Observatory (SDO)~\citep{SDO}. The first row shows the SUIT images with the bounding boxes inferred by rescaling each image by subtracting the minimum and dividing by the range of intensities, as is the standard scheme of rescaling for image data. For such an image, we find the that the model fails to capture many of the small-scale structure, especially the sunspots, while comparing the estimated sunspots with the AIA data. We perform a trial and error of rescaling, and find that a rescaling of each image by a factor of $8$ helps in better identification of the features in the SUIT images. These estimated boxes are shown in the second row of Figure~\ref{fig:SUIT_DATA}.

First off, we find that the SUIT observations show several features of interest. The model is able to capture several plage regions very well with high confidence -- for instance, the big plage in the southern hemisphere in the left column of Figure~\ref{fig:SUIT_DATA} is captured with a confidence of $>0.8$. Second, we see several sunspots being captured by the model. Interestingly the model seems to have a wide range of confidence levels in capturing these sunspots. This may occur due to a significant perceptual contrast difference between the mock and real SUIT images. In fact, we find the large plage region in the northern hemisphere of the left panel is not captured by the model -- this effect could arise from the contrast, and a lack of such a strongly oriented feature in the training set. We note that the level 1.1 SUIT data is scattered light and flat field corrected, and have been recorded with an exposure time of 1.4 seconds. However, there are variations in intensity over time as the net throughput changes due to contaminant deposition on the CCD. This gets cleaned after every baking cycle, until it gets dim again. This inconsistency of throughput is not captured by the training set and the model, and hence a manual normalization is necessary at the moment.

\subsection{Evaluation Using Statistical and Tamura Measures}
We now perform a comparative analysis of the \textit{Feature parameters}. The counts of these individual features, both ground truth and predicted, are given in Table~\ref{tab:events}. These regions are chopped out, using the detected coordinates from the original FITS data and not the rescaled data, to ensure that we capture the non-altered values. Hence, some of the numbers may not be precisely comparable between the mock-SUIT and SUIT data. We note that the number of features is different between ground truth and predicted as the manually labeling process may miss a few regions, which are however captured by the model. The number of features in the test set are extracted by deploying the final weights on the SUIT data.
\begin{table}[!ht]
    \centering
    \begin{tabular}{llllllllllllllllll}
    \hline
         & Plages & Sunspots & Filaments & Off-limb \\
    \hline
        Training Set (Ground Truth) & 407 & 246 & 126 & 364 \\
        Training Set (Predicted) &564 & 310 & 161 & 237\\
        Validation Set (Ground Truth) & 98 & 45 & 20  & 164\\
        Validation Set (Predicted) & 133 & 64 & 25  & 124\\
        Test(SUIT Data) & 420 & 251& NA & 42\\
    \end{tabular}
   \caption{Number of ground truth and detected regions samples chopped out from the full disc images.}
    \label{tab:events}
    
\end{table}

\begin{figure}[ht!]
    \centering
    \includegraphics[width=\linewidth]{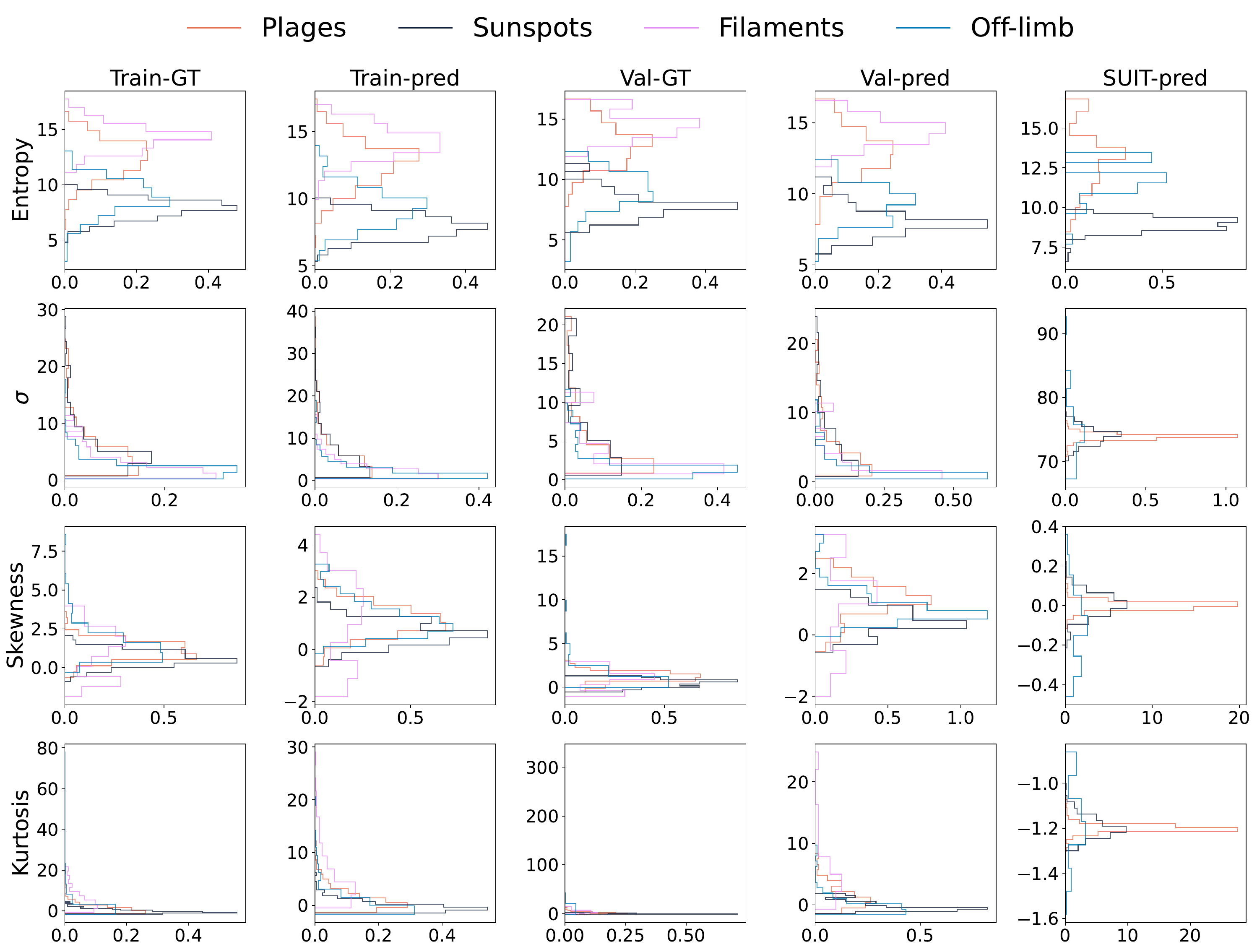}
    \caption{Variation of the statistical measures for the features of interest, in terms of entropy, $\sigma$, skew and kurtosis along each \textit{row}. The columns correspond to the train ground truth, train prediction, validation ground truth, validation prediction, and SUIT prediction respectively. Sunspots are marked with \textit{dark-gray color}, plages with \textit{yellow-orange color}, filaments with \textit{pink} and off-limb regions with \textit{blue color}.}
    \label{fig:stats}
\end{figure}

\begin{figure}[ht!]
    \centering
    \includegraphics[width=\linewidth]{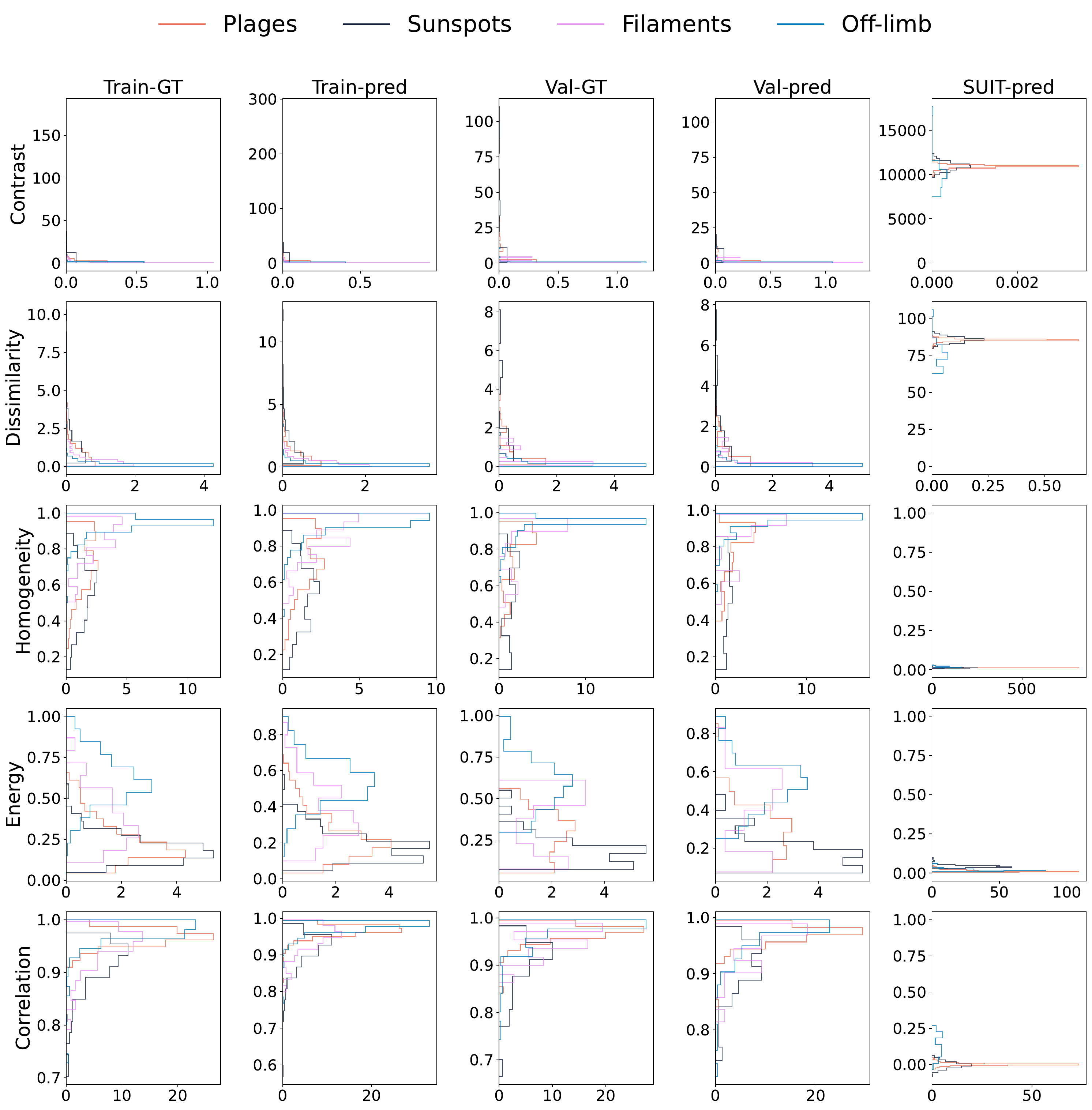}
    \caption{Variation of the Tamura measures for the features of interest, including contrast, dissimilarity, homogeneity, energy, and correlation. The \textit{columns} correspond to the train ground truth, train prediction, validation ground truth, validation prediction, and SUIT prediction respectively. Sunspots are marked with \textit{dark-gray color}, plages with \textit{yellow-orange color}, filaments with \textit{pink} and off-limb regions with \textit{ blue color}. }
    \label{fig:tamura}
\end{figure}

In Figure~\ref{fig:stats}, we present the distribution of Entropy, $\sigma$, skewness, and kurtosis computed within each bounding box for the training set (GT and prediction), validation set (GT and prediction), and SUIT data. The different colors correspond to different features of interest.

From the first row of Figure~\ref{fig:stats}, we find that entropy clearly differentiates between the four features. We find that distributions for predictions align fairly well with the ground truth for both training and validation sets. We find that the sunspots have a lower entropy than plages, which have lower entropy than filaments. The sunspots in our bounding boxes are seen with almost similar intensities in the box, while in plages, we see far more structure and intensity variation. The filament bounding boxes encompass a large area, with most of the bounding box consisting of non-filamentary structures with varied intensities. Since entropy is a measure of the number of non-unique intensities in each feature, we find the entropy distributions in the increasing order of sunspots, plages and filaments in Figure~\ref{fig:stats} first row. We note that the off-limb structures lie somewhere between sunspots and plages in the dataset. However, the SUIT-pred panel shows a notable increase in the location of the peak of the distribution for plages and off-limb regions, suggesting some inconsistency in model predictions on real SUIT data.

From the second row of Figure~\ref{fig:stats}, we find that the standard deviation $\sigma$ spans a wide range across all data. Predictions for standard deviation appear relatively consistent with the ground truth across training and validation sets. The distribution of $\sigma$ however changes in the SUIT data, with all the distributions being more consistent with each other. If we were to consider only the median values of the distribution, they are mostly consistent across the all of the datasets. Thus, standard deviation does not seem to be a very good discriminator of the selected features, and does not aid us in an auto-validation procedure.

In the third row of Figure~\ref{fig:stats}, we present the skewness of the intensities in different bounding boxes. In the training set, we find the filament intensity skewness to span a wide range, while the other features show a distribution peaked near $0$, but on the positive side. This is also noticed in the prediction from the training set. In the validation set, we find the skew to be mostly positive, with the predictions consistent with the ground truth. The SUIT skew are similarly peaked near $0$, but we find that the off- limb structures have a large spread of skew values unlike the mock dataset. The intensity distribution in any of the features of interest generally has a long tail towards the higher end than the lower end, and this is captured by the skewness. However, we note that the skewness of intensity distribution is again not a good discriminator of the selected features, and does not aid us in an auto-validation procedure.

In the final row of Figure~\ref{fig:stats}, we present the kurtosis of the intensity distribution in the different bounding boxes. In general, we find a good agreement in kurtosis distributions between predictions and ground truth for training and validation data. We find that the filaments and plages tend to have more samples with larger kurtosis, indicating a distribution falling off faster than a Gaussian. This seems to reflect a very long tail of intensities, which is also consistent with the higher entropy of these regions from the first row of this Figure However, the kurtosis for SUIT-pred data exhibits tighter and sharper peaks, while having the opposite sign. This indicates a potential underestimation of tails in intensity distributions, particularly in off-limb areas. Furthermore, this is also reflective of the difference in the mock-SUIT and real-SUIT observations, which are seen as inconsistencies here.

Overall, we find that the model shows reasonable performance in replicating training and validation data distributions. While the actual values of these statistical measures are different with SUIT data, the distributions are mostly consistent across mock SUIT and SUIT predictions. We note that the divergence of results occurs especially in the off-limb regions.

In Figure ~\ref{fig:tamura}, we present the distributions of contrast, dissimilarity, homogeneity, energy, and correlation measures for each bounding box, as described in Section~\ref{sec:STF}. These are computed within each bounding box for the training set (GT and prediction), validation set (GT and prediction), and SUIT data. The different colors correspond to different features of interest, given here as sunspots (dark gray), plages (yellow-orange), filaments (pink), and off-limb regions (blue).

In the first row of Figure~\ref{fig:tamura}, we display the contrast across the different features. The ground truth contrast distributions for training and validation data show relatively low contrast values across structures, especially for plages, sunspots, and off-limb regions. The model captures this in training and validation predictions, though there is some spread in SUIT-pred. However, we note that the contrast values are larger in sunspots, when compared to plage and filament regions. Interestingly, the absolute contrast values are 2 orders of magnitude larger in the SUIT images
,which point to differences in the actual values of the mock and real SUIT data. We also note that the distribution in SUIT boxes itself  in quite symmetric, unlike the mock data.

In the second row of Figure~\ref{fig:tamura}, we display the dissimilarity for the different features of the datasets. We again find a distinction in the distributions for mock-SUIT and SUIT data. We find the dissimilarity in training and validation data are generally well-aligned between ground truth and predictions. In the mock-SUIT data, we find sunspots to have higher dissimilarity over plage and filament regions. However, sunspots also have low entropy from the first row of Figure~\ref{fig:stats}. This tells us that the dissimilarity is large in sunspots due to an emphasis of the metric on boundaries. The core structures themselves, are, however of similar intensities, which results in a lower entropy of the first row of Figure~\ref{fig:stats}.

In the third row of Figure~\ref{fig:tamura}, we display the homogeneity of the for different features. We find each feature to present a distinct distribution across all data. We see that sunspots are the least homogeneous, followed by plage, off-limb structures, and filaments. The absolute homogeneity numbers are however less for SUIT observations. The model distribution from predictions well-match the ground truth. Homogeneity, while not complete, can be used as a measure to differentiate between these features.

In the fourth row of Figure~\ref{fig:tamura}, we display the energy of the images. Essentially, the energy reflects the concentration of certain intensity values. We find off-limb and filaments to have higher energy. However, this is flipped in the SUIT images, with sunspots showing as an outlier distribution. Furthermore,  SUIT-pred shows substantially lower energy values across all structures, particularly off-limb areas.  

Finally, we present the correlation measure in the last row of Figure~\ref{fig:tamura}. We find the features to have similar autocorrelation datasets in mock-SUIT, and this changes in SUIT data. The correlation values are quite high for all structures in training and validation data, indicating a consistent spatial relationship between neighboring pixels. Predictions align well with the ground truth for these datasets. SUIT-pred, however, exhibits lower correlation values, particularly for off-limb regions, suggesting that the model may be struggling with accurately predicting spatial relationships in Mg II k passband images. 

Overall, we find a good consistency between the \textit{feature properties} of the GT and the predicted bounding boxes for the mock SUIT data. Furthermore, we also find a good consistency between the mock SUIT and SUIT data in several of the \textit{feature properties}. 

\section{Discussion and Conclusions}
In this work, we have developed {\spc}, an event detection system for SUIT. Specifically, we develop a system to identify sunspots, filaments, plage and off-limb regions. This framework is based on a YOLO v8-obb model, and is trained on mock-SUIT data generated from IRIS mosaics. We perform a comparison across different YOLO versions, and select the best performing model for our feature identification purposes. The best model achieves a precision of 0.85, recall of 0.99, and an MAP of 0.97 and 0.90 on JPG files and a  precision of 0.79, recall of 0.86, and an MAP of 0.87 and 0.82 for FITS data respectively. 

We deploy this system on SUIT data, and find that the model picks up most of the regions of interest, if we perform a visual inspection. However, since the GT is generated by hand-labeling individual features of interest, the system lacks a more objective measure of determining model performance on actual SUIT images in the absence of a ground truth dataset. 

To this end, we develop a self-validation scheme based on statistical and Tamura measures, which provide a comparison of these measures across different features of interest. These features are based on the statistical moments and gray level co-occurrence matrix for the intensity values for each feature, described in Section~\ref{sec:STF}. The proposed metrics capture various properties of the intensity distributions for our features of interest. A statistical distribution of these metrics for the GT and predicted bounding boxes for different features, along with the metrics for SUIT data would enable an automatic self-validation of the performance of the model. We perform this analysis and note the results in Section~\ref{sec:results}.

We find entropy to be the best metric statistically differentiating the features of interest. The distribution of entropy for all samples follows similar trends across both the ground truth and predictions of mock-SUIT and SUIT data. This gives us confidence in the performance of our model on unseen data, and lets us perform a zero-shot learning of SUIT data. 

Among the other metrics considered, we find contrast, dissimilarity, homogeneity and energy to also provide an understanding of the differences between the features. While the distribution of these metrics have a great overlap across the different features, some of the features have strong signals that help us understand the consistency across all dataset. 

We find that sunspots exhibit low entropy, high dissimilarity and contrast, and low homogeneity and correlation across all datasets. Since sunspots are structures with uniform intensity, there are not many `unique' pixel intensities, and this results in a relatively lower entropy when compared to other features. However, this is in tension with high dissimilarity and contrast, which are also measures of the number of non-unique pixel intensities in the region of interest. The sunspots, due to sharp intensity changes at the boundary, result in a high dissimilarity and contrast, but lower entropy. We note that while the energy is generally consistent with other features, the SUIT data and the validation GT shows sunspots with high energy, corresponding to several similar intensity values, reflecting the entropy of the intensities. Finally, the low homogeneity and correlation are possibly the effect of the boundaries, which may wrongly lead us to interpret sunspots as regions with a lack of spatial uniformity. 

We find that the filaments display high entropy and mild dissimilarity, low contrast, medium homogeneity and correlation, high positive and negative skewness, and high kurtosis. The filaments are long structures, with their bounding boxes encompassing both the filament and surrounding regions. This highlights their complex, varied textures and uneven intensity transitions across the structure, and hence are seen as features with high entropy. Since the filaments in mock-SUIT data are seen very weakly, we see this structures with lower contrast than plage or sunspot regions, reflecting more gradual intensity changes without sharp boundaries. Since the filament regions contain a lot of background + the filament, these regions have low homogeneity and correlation, indicating inconsistent intensity patterns and less predictable pixel relationships across the structure. We note the rather high skewness and kurtosis for filaments arise from the different kinds of background they can have, resulting in intensity distributions with a long tail or sharp peaks.

Plages are regions with relatively uniform, and bright intensities. These are captured by our metrics as low entropy, low dissimilarity, high homogeneity, and high correlation. They also have high contrast which reflects subtle variations in brightness within plages rather than abrupt transitions. They also show moderate skewness and kurtosis, indicating a somewhat peaked but symmetrical intensity distribution. 

The off-limb regions are defined as structures seen outside the disc of the Sun. These regions could be any intensity structures, but are most probably prominences. These off-limb regions have low entropy, low contrast, and high homogeneity, capturing their relatively empty and uniform background intensity. Low dissimilarity and correlation indicate consistent intensity patterns without structured variations. They also have very low skewness, that reflect the uniform, low-intensity nature of these regions, with minimal brightness variation.

The proposed auto-validation scheme evaluates the extracted correctness of the bounding boxes. It does not, however, provide a measure of the fraction of all regions captured correctly by the model. For this purposes, we will again to need to resort to manual labeling of data. 

In summary, we find that we can explain the behavior of these measures with our features of interest, and this behavior is consistent with known observations. With such measures of comparison, we find that the  model performs well in capturing statistical and Tamura measures for the training and validation datasets. However, many of the numbers are very different from the actual SUIT data, which is likely due to the discrepancy between the scheme for mock-SUIT image generation, and the SUIT images themselves. We note that a deployment of this model would need fine-tuning on real SUIT data for a more effective identification of features. We plan to perform a semi-supervised training of the {\spc} model using the labeled SUIT data with a human-in-loop system, while having a continuous learning framework to account for changes in SUIT data over time. Adjusting the model to better account for these differences will improve generalization to SUIT data, especially for complex structures like filaments and off-limb regions.
\\

\noindent\textbf{Acknowledgments} 
P.S. would like to thank the Inter-University Centre for Astronomy and Astrophysics, Pune for providing support during his research visit. U.V and D.T would like to acknowledge funding from ISRO/RESPOND for the project ``Solar Flares: Physics and Forecasting for Better Understanding of Space Weather'', ISRO/RES/2/438/22-23. U.V. and D.T. would like to acknowledge the NVIDIA Academic Hardware grant provided to U.V. and IUCAA. U.V. would like to thank Carlos Diaz Baso for comments and suggestions on the work and manuscript. All the authors would like to thank the reviewer for their comments and suggestions.
    
    This research used Numpy~\citep{numpy}, Scikit-image~\citep{scikit-image}, Open-CV~\citep{opencv_library}.pillow~\citep{clark2015pillow}, pytorch~\citep{NEURIPS2019_9015}, torchvision~\citep{torchvision}, seaborn~\citep{michael_waskom_2017_883859}, ultralytics~\citep{yolov8_ultralytics,v9}, astropy~\citep{price2018astropy}, scipy~\citep{2020SciPy-NMeth}, sunpy~\citep{sunpy}, matplotlib~\citep{hunter2007matplotlib} python libraries.

\begin{fundinginformation}
    Aditya-L1 is an observatory class mission which is funded and operated by the Indian Space Research Organization. The mission was conceived and realized with the help from all ISRO Centres and payloads were realised by the payload PI Institutes in close collaboration with ISRO and many other national institutes - Indian Institute of Astrophysics (IIA); Inter-University Centre of Astronomy and Astrophysics (IUCAA); Laboratory for Electro-optics System (LEOS) of ISRO; Physical Research Laboratory (PRL); U R Rao Satellite Centre of ISRO; Vikram Sarabhai Space Centre (VSSC) of ISRO. SUIT is built by a consortium led by the Inter-University Centre for Astronomy and Astrophysics (IUCAA), Pune, and supported by ISRO as part of the Aditya-L1 mission. The consortium consists of SAG/URSC, MAHE, CESSI-IISER Kolkata (MoE), IIA, MPS, USO/PRL, and Tezpur University.  IRIS is a NASA small explorer mission developed and operated by LMSAL with mission operations executed at NASA Ames Research Center and major contributions to downlink communications funded by ESA and the Norwegian Space Centre.
\end{fundinginformation}

\begin{ethics}
\begin{conflict}
The authors declare no competing interests.\\
\end{conflict}
\end{ethics}

\bibliographystyle{spr-mp-sola}
\bibliography{references}

\end{document}